\documentclass[letterpaper,english,reprint, pra]{revtex4-2}
\usepackage[T1]{fontenc}
\usepackage[latin9]{inputenc}
\usepackage{color}
\usepackage{babel}
\usepackage{verbatim}
\usepackage{mathrsfs}
\usepackage{amsmath}
\usepackage{amssymb}
\usepackage{graphicx}
\usepackage{esint}
\usepackage{xargs}[2008/03/08]
\usepackage[unicode=true,pdfusetitle,
 bookmarks=true,bookmarksnumbered=false,bookmarksopen=false,
 breaklinks=false,pdfborder={0 0 0},pdfborderstyle={},backref=false,colorlinks=true]
 {hyperref}

\makeatletter

\usepackage{orcidlink}
\usepackage{graphicx}

\usepackage{tikz}
\usetikzlibrary{graphs, graphs.standard}

\makeatother

\begin{document}
\title{Quantum control for Zeno effect with noise}
\author{Haorui Chen$^{1}$\orcidlink{0009-0006-3520-4240}}
\author{Shengshi Pang$^{1,2}$\orcidlink{0000-0002-6351-539X}}
\email{pangshsh@mail.sysu.edu.cn}

\affiliation{$^{1}$School of Physics, Sun Yat-sen University, Guangzhou, Guangdong
510275, China}
\affiliation{$^{2}$Hefei National Laboratory, University of Science and Technology
of China, Hefei 230088, China}
\date{\today}
\begin{abstract}
The quantum Zeno effect is a distinctive phenomenon in quantum mechanics, describing the nontrivial effect of frequent projective measurements on hindering the evolution of a quantum system. However, when subjected to environmental noise, the quantum system may dissipate, and the quantum Zeno effect no longer works. This research starts from the physical mechanism for the decay of the quantum Zeno effect in the presence of noise and investigates the effect of coherent quantum controls on mitigating the decrease of the survival probability that the system stays in the initial state induced by the noise. We derive the decay rate of the survival probability with and without coherent quantum controls in general, and show that when the frequency of the projective measurements is large but finite, proper coherent controls by sufficiently strong Hamiltonians can be designed to decrease the decay rate of the survival probability. A two-level quantum system suffering from typical unitary and nonunitary noise is then considered to demonstrate the effect of the proposed coherent quantum control scheme in protecting the quantum Zeno effect against the noise. The decay rate of the survival probability is obtained in the presence of noise, and the control Hamiltonian is further optimized analytically to minimize the decay rate by a variational approach. The evolution paths of the quantum system with the optimal coherent controls are illustrated numerically for different scenarios to explicitly show how the coherent control scheme works in lowering the decay of survival probability.
\end{abstract}
\maketitle
\global\long\def\huaD{\mathscr{\mathcal{D}}}%
\newcommandx\huaL[1][usedefault, addprefix=\global, 1=]{\mathcal{L}_{\mathrm{#1}}}%
\newcommandx\rig[1][usedefault, addprefix=\global, 1=]{|#1\rangle}%
\newcommandx\lef[1][usedefault, addprefix=\global, 1=]{\langle#1|}%
\global\long\def\psio{\psi_{0}}%
\global\long\def\gammaeff{\gamma_{\mathrm{eff}}}%
\global\long\def\noise{\mathrm{n}}%
\global\long\def\control{\mathrm{c}}%
\global\long\def\optimal{\mathrm{opt}}%
\global\long\def\kmin{k_{\mathrm{min}}}%
\global\long\def\gammafree{\gamma_{\mathrm{free}}}%
\global\long\def\tr{\mathrm{Tr}}%
\global\long\def\me{\mathcal{E}}%
\newcommandx\var[1][usedefault, addprefix=\global, 1=H]{\langle\Delta^{2}#1\rangle}%
\global\long\def\abs#1{\left|#1\right|}%
\newcommandx\pt[1][usedefault, addprefix=\global, 1=]{\partial_{\tau}^{#1}}%
\global\long\def\dlim#1{\left.#1\right|_{\tau=0}}%
\global\long\def\vp{\boldsymbol{\sigma}}%
\global\long\def\im{{\rm Im}}%
\global\long\def\zhuan{\mathrm{T}}%
\global\long\def\zheng#1{\mathrm{#1}}%
\global\long\def\inten{\mu}%
\newcommandx\prot[1][usedefault, addprefix=\global, 1=]{p_{\mathrm{#1}}}%
\newcommandx\hami[1][usedefault, addprefix=\global, 1=]{H_{\mathrm{#1}}}%
\newcommandx\rou[1][usedefault, addprefix=\global, 1=]{\rho_{\mathrm{#1}}}%
\newcommandx\thetakong[1][usedefault, addprefix=\global, 1=]{\theta_{\mathrm{#1}}}%
\newcommandx\ph[1][usedefault, addprefix=\global, 1=]{\phi_{\mathrm{#1}}}%
\global\long\def\tot{{\rm tot}}%
\global\long\def\ddlim#1{\left.#1\right|_{\tau=0}}%
\newcommandx\huaLh[1][usedefault, addprefix=\global, 1=]{\mathcal{L}_{H_{\mathrm{#1}}}}%
\global\long\def\rot{\rho_{t}}%
\global\long\def\nb{\boldsymbol{\nu}}%
\global\long\def\ro{\boldsymbol{r_{0}}}%

\section{Introduction}

The quantum Zeno effect is the quantum version of the classical Zeno
effect, initially proposed by the ancient Greek philosopher Zeno,
who is known for the famous paradoxes such as the \textquotedblleft flying
arrow\textquotedblright{} and \textquotedblleft Achilles and the tortoise\textquotedblright{}
\citep{aristotle}. Interestingly, while the Zeno effect is a paradox
in the classical world hypothesizing that frequent observation can
freeze the evolution of a system, which is certainly not possible
in real life, the capability of quantum measurements to project quantum
systems onto specific states \citep{neumann1932mathematische} opens
up the possibility for realizing the Zeno effect in the quantum realm.
As early as 1967, Beskow and Nilsson observed that frequent measurements
of the positions of unstable particles in a cloud chamber effectively
prevented the decay of the particles\citep{beskow1967concept}. This
discovery sparked widespread interest among physicists and mathematicians
in the feasibility of the Zeno effect in quantum mechanics, leading
to subsequent confirmations of the quantum Zeno effect with different
experimental setups and physical systems \citep{itano1990quantum,kwiat1995interactionfree,nagels1997quantum,streed2006continuous,bernu2008freezing,signoles2014confined,schafer2014experimental,slichter2016quantum}
and extensive intriguing theoretical explorations \citep{exner2005aproduct,exner2007zenoproduct,burgarth2019generalized,exner2021noteon}.

The standard mechanism for realizing the quantum Zeno effect is to
freeze the evolution of a quantum system through frequent projective
measurements \citep{misra1976thezenotextquoterights,koshino2005quantum,facchi2008quantum},
similar to the phenomenon in the classical Zeno effect known by the
old saying \textquotedblleft A watched pot never boils.\textquotedblright{}
With further research, Kofman found that the anti-Zeno effect, accelerating
the evolution of a quantum system proposed by Kaulakys and Gontis
in 1997 in the context of quantum chaos \citep{kaulakys1997quantum},
would be a more common phenomenon in the quantum regime \citep{kofman2000acceleration},
in contrast to the quantum Zeno effect. This has made the relation
and crossover between quantum Zeno and anti-Zeno effects a hot topic
in quantum mechanics \citep{facchi2001fromthe,koshino2005quantum,facchi2008quantum}.
Moreover, the quantum Zeno effect has been generalized to the quantum
Zeno dynamics through performing frequent projective measurements
on a proper subspace of a quantum system known as the Zeno subspace,
where nontrivial unitary evolution is allowed inside the Zeno subspace
while the evolution outside the Zeno subspace is suppressed \citep{facchi2002quantum,facchi2008quantum,raimond2010phasespace,paz-silva2012zenoeffect}.

Currently, various approaches to the quantum Zeno effect have been
proposed. Based on the characteristic timescales of quantum operations
that realize the quantum Zeno effects compared to the timescales of
quantum system free evolutions, the quantum Zeno effects can be broadly
categorized into pulsed quantum Zeno effects and continuous quantum
Zeno effects \citep{facchi2008quantum,burgarth2020quantum,streed2006continuous}.
The pulsed quantum Zeno effects are realized through frequent projective
measurements as mentioned above or strong unitary operations (often
known as unitary kicks), which can be unified with bang-bang control
and dynamical decoupling in suppressing the decoherence of open quantum
systems \citep{viola1998dynamical,viola1999dynamical,duan1999suppressing},
both equivalent in the Zeno limit \citep{facchi2004unification}.
The continuous quantum Zeno effect describes the quantum Zeno effect
induced by continuous strong coupling between the main and ancillary
systems \citep{schulman1998continuous,facchi2002quantum}, by large
dissipations that leads the quantum system to decay into a stable
subspace \citep{nakazato1999twolevel,macieszczak2016towards}, by
continuous partial measurements\citep{snizhko2020quantum,kumari2023qubitcontrol}
or by non-selective continuous measurements \citep{ruskov2006quantum,gherardini2016stochastic,kumar2020quantum,elattari2000effectof,gurvitz2003relaxation,Presilla1996}.
In recent years, intensive research has been dedicated to the connections
and unified theoretical frameworks between different manifestations
of the quantum Zeno effects \citep{facchi2002quantum,facchi2004unification,facchi2008quantum,facchi2005control,burgarth2020quantum,burgarth2019generalized2,hahn2022unification}.
At the same time, attempts to explore the competition between different
methods simultaneously applied in the quantum Zeno effects, e.g.,
involving both nonselective continuous measurements and large dissipation,
has started to emerge \citep{kumar2020quantum}.

In analogy to many other quantum effects, an essential ingredient
to realize the quantum Zeno effect is the coherence of the quantum
system, which ensures the probability that the system stays in the
initial state decays quadratically with time in a short time interval.
However, practical quantum systems are inevitably disturbed by the
noise from the environments, and quantum coherence is vulnerable to
the detrimental effects such as decoherence, relaxations, and dissipations
\citep{blanchard2000decoherence,braun2013dissipative} which can spoil
the quantum Zeno effects and quantum Zeno dynamics \citep{gurvitz2003quantum,gurvitz2003relaxation,guerra2012decoherence,kumar2020quantum}.
To protect quantum systems against the noise, quantum techniques such
as decoherence-free subspaces \citep{duan1998reducing,shabani2005theoryof,lidar2003decoherencefree2},
coherent control schemes \citep{peirce1988optimal,judson1992teaching}
and quantum error correction codes \citep{shor1995schemefor,steane1996errorcorrecting}
have been developed, and in fact, the quantum Zeno effect is essential
to some quantum error correction techniques \citep{vaidman1996errorprevention,duan1998prevention,yang2002errorprevention,facchi2004unification,erez2004correcting}
underscoring its significance in the realm of quantum information
science. The quantum Zeno effect, including the quantum Zeno dynamics,
has found versatile applications due to its simplicity and diversity
in realization, ranging from realization of decoherence-free subspaces
for quantum gates \citep{beige2000quantum} to utilization of classical
noise and engineering of non-Markovianity in quantum simulation \citep{stannigel2014constrained,patsch2020simulation},
diagnosis of noise correlations between photon polarizations \citep{virz`i2022quantum},
realization of universal quantum control between noninteracting qubits
\citep{blumenthal2022demonstration}, and optimization of quantum
algorithms \citep{herman2023constrained}, etc.

The reservoir correlation time is critical to the effect of noise
on the quantum Zeno effect. For example, Gurvitz \citep{gurvitz2003relaxation}
found that quantum system can still be frozen if the reservoir correlation
time is finite, i.e., the noise is non-Markovian, while the Zeno effect
vanishes in the short correlation limit, i.e., the noise is Markovian.
In recent years, there has been an increasing interest in research
devoted to the quantum Zeno effect in the presence of large Markovian
dissipations \citep{popkov2018effective,kumar2020quantum,popkov2021fullspectrum}.
For instance, Popkov \emph{et al.} derived that the effect of strong
local dissipation in the Zeno limit is equivalent to Markovian quantum
dynamics featuring a renormalized effective Hamiltonian and weak dissipation.

As Markovian noise can spoil the quantum Zeno effect and the survival
probability of the initial state decays exponentially with time in
the presence of noise, it is an intriguing question whether it is
possible and how to decrease the noiseinduced decay of the quantum
Zeno effect by modulating the dynamics of the quantum system.

In this work, we study these questions in detail by involving Markovian
noise in the dynamics of a quantum system. The influence of the noise
on the quantum Zeno effects and the decay of quantum systems with
noise in the Zeno limit are investigated in general, revealing the
potential for decreasing the decay rate of survival probability that
the system stays in the initial state by quantum controls. We consider
controls on the Hamiltonian of the system to protect the Zeno effect
against the noise in this paper, and show that the Hamiltonian control
needs to be strong with a strength proportional to the measurement
frequency, which is large but finite in order to decrease the influence
of noise on the quantum Zeno effect. We obtain the decay rate of the
survival probability in the presence of noise with strong control
Hamiltonians in general, and show the conditions on the control Hamiltonian
as well as on the frequency of the projective measurements to mitigate
the disruption on the quantum Zeno effect caused by the noise. This
Hamiltonian control scheme is then applied to a two-level system with
typical unitary and nonunitary noise to illustrate the general results.
We consider the dephasing and the amplitude damping noise as examples,
and obtain the minimum decay rate of the survival probability by optimizing
the Hamiltonian controls. The results show that the survival probabilities
of the initial state can indeed be increased by the optimized Hamiltonian
controls on the quantum system. The evolution paths of the two-level
system engineered by the Hamiltonian controls are visualized on the
Bloch sphere by numerical simulations to illustrate how the control
scheme protects the survival probability against the two types of
noise.

The paper is organized as follows. In Sec.$\ $\ref{sec:PRELIMINARIES},
we provide preliminaries for the theory of open quantum systems and
the quantum Zeno and anti-Zeno effects. In Sec.$\ $\ref{sec:STRONG-QUANTUM-CONTROL},
we study the decay of quantum Zeno effect in the presence of Markovian
noise, and derive the effective decay rate of the survival probability
in the presence of Hamiltonian control, which further shows the conditions
on the control Hamiltonian to reduce the decay of the survival probability
caused by Markovian noise. Sec.$\ $\ref{sec:STRONG-QUANTUM-CONTROL-1}
considers a strong Hamiltonian control scheme for a two-level system
in the presence of two different types of noise, and obtain the effective
decay rate and the optimal Hamiltonian controls for the two types
of noise respectively. The optimal control Hamiltonians are further
derived analytically, and the physical mechanism for the optimal coherent
control schemes to suppress the noise influence on the survival probability
is illustrated numerically and analyzed in detail. The paper is finally
concluded in Sec.$\ $\ref{sec:CONCLUSION}.

\section{Preliminaries\label{sec:PRELIMINARIES}}

In this section, we briefly introduce the preliminary knowledge of
the open quantum system theory and the quantum Zeno effect relevant
to the current research.

\subsection{Dynamics of open quantum systems\label{subsec:Quantum-operations}}

In a closed quantum system, the evolution of a quantum state is generally
described a unitary transformation,
\begin{eqnarray}
\me\left[\rho\right] & = & U\rho U^{\dagger},\label{eq:1}
\end{eqnarray}
where $U$ is the unitary evolution operator,
\begin{equation}
U=\exp(-iHt),
\end{equation}
determined by the Hamiltonian $H$ of the system and the evolution
time $t$. However, for an open quantum system exposed to the environment,
the dynamics of the system can no longer be described by unitary evolutions
because of the inevitable coupling between the system and the environment.

For an open quantum system, by treating the system and environment
as a closed joint system, the total Hamiltonian of the system and
the environment can be written as
\begin{equation}
\hami[tot]=\hami[S]+\hami[E]+\hami[SE],
\end{equation}
where $\hami[S]$ and $\hami[E]$ are the local Hamiltonians which
rule the dynamics of system and environment, respectively, and $\hami[SE]$
stands for the interaction Hamiltonian between the system and the
environment.

uppose that the system and the environment are initially uncorrelated.
The initial joint state of the system and the environment can be written
as $\rou[SE]=\rou[S]\otimes\rou[E]$, where $\rou[S]$ and $\rou[E]$
are the density operators of the system and the environment, respectively,
and the unitary evolution of the joint state can be written as
\begin{equation}
\me_{\left(t,0\right)}\left[\rou[SE]\right]=U\left(t\right)\left(\rou[S]\otimes\rou[E]\right)U^{\dagger}\left(t\right).
\end{equation}

When one is interested in the system only, the joint evolution of
the system and environment can be reduced to the system alone by tracing
over the degrees of the freedom of the environment,
\begin{equation}
\me_{\left(t,0\right)}\left[\rou[S]\right]=\tr_{\zheng E}\left[U\left(t\right)\left(\rou[S]\otimes\rou[E]\right)U^{\dagger}\left(t\right)\right].\label{eq:eto}
\end{equation}

The quantum evolution $\mathcal{E}_{\left(t,0\right)}$ obtained in
Eq. \eqref{eq:eto} gives the general dynamical process of an open
quantum system coupled to the environment.

An important property of a quantum process is the Markovianity based
on the completely positive and trace-preserving (CPTP) divisibility
of the process. If a quantum process satisfies the CPTP divisibility
condition,
\begin{equation}
\me_{\left(t_{n},t_{0}\right)}=\me_{\left(t_{n},t_{n-1}\right)}\me_{\left(t_{n-1},t_{n-2}\right)}\cdots\me_{\left(t_{1},t_{0}\right)},
\end{equation}
where $t_{n}\geq t_{n-1}\geq\cdots\geq t_{0}$ are arbitrary time
points and each $\me\left(t_{k+1},t_{k}\right)$ is a CPTP quantum
map, the quantum process $\me\left(t_{n},t_{0}\right)$ is called
Markovian, otherwise non-Markovian. The Markovianity of quantum dynamics
is closely related to the reservoir correlation time, which determines
the memory effects of the environment, and dependent on various ingredients
such as the dimension of the environment and the strength of interaction
between the system and environment \citep{lindblad1976onthe,gorini2008completely,breuer2007thetheory}.

According to the open quantum system theory that the Markovian dynamics
of an open quantum system can always be described by a Gorini-Kossakowski-Lindblad-Sudarshan
master equation \citep{2007quantum,chruscinski2017abrief},
\begin{equation}
\frac{d\rho\left(t\right)}{dt}=\huaL[\mathit{t}]\left[\rho\left(t\right)\right]=-i\hbar\left[H,\rho\left(t\right)\right]+\sum_{k}\inten_{k}\left(t\right)\huaD[V_{k}]\rho\left(t\right),\label{eq:mastereq}
\end{equation}
where $H$ is the Hamiltonian of the system and $\huaD[V_{k}]$ denotes
the Lindblad infinitesimal generator for dissipative process induced
by the $k$th noise channel generally in the form
\begin{equation}
\huaD[V_{k}]=V_{k}(\cdot)V_{k}^{\dagger}-\frac{1}{2}\left\{ V_{k}^{\dagger}V_{k},\cdot\right\} ,\label{eq:3}
\end{equation}
with the Born-Markov approximation \citep{breuer2007thetheory}, where
$\left[\cdot,\cdot\right]$ and $\left\{ \cdot,\cdot\right\} $ denote
the commutator and the anticommutator, respectively. It can be proven
that a quantum process is Markovian if and only if it can be described
by a master equation \eqref{eq:mastereq} with all coefficients $\inten_{k}\left(t\right)$'s
non-negative for any time $t$ \citep{rivas2012openquantum}. When
Markovian noise are considered in the following sections, we use the
master equation \eqref{eq:mastereq} to involve the noise in the evolution
of the quantum system.

\subsection{Zeno and anti-Zeno effect\label{subsec:Zeno-and-anti-Zeno}}

In this subsection we will briefly introduce the fundamental knowledge
about the quantum Zeno and anti-Zeno effects.

Suppose a closed quantum system ruled by a Hamiltonian $H$ is initially
prepared in a pure state $|\psi\rangle$. One can perform a projective
measurement after an evolution time $t$ of the system to verify whether
the system is still in its initial state, and the survival probability
is given by
\begin{equation}
p\left(t\right)=|\langle\psi|e^{-iHt}|\psi\rangle|^{2}.\label{eq:53}
\end{equation}
If the projective measurement is carried out repetitively at time
interval $\tau$ during an evolution time $t$, the final survival
probability of the system in the initial state at time $t$ reads

\begin{equation}
P\left(t\right)=p\left(\tau\right)^{t/\tau},\label{eq:pt}
\end{equation}
which can be rewritten as an exponential decay with time $t$,
\begin{equation}
P\left(t\right)=\exp\left(-\gammaeff\left(\tau\right)t\right),\label{eq:27}
\end{equation}
and $\gammaeff\left(\tau\right)$ is the effective decay rate given
by
\begin{equation}
\gammaeff\left(\tau\right)=-\frac{\ln p\left(\tau\right)}{\tau}.\label{eq:11}
\end{equation}

If the interval $\tau$ between two consecutive measurements is short,
the probability $p(\tau)$ can be approximated to the second order
of $\tau$,
\begin{equation}
p(\tau)\approx1-\tau^{2}\var,\label{eq:ptau}
\end{equation}
where $\var=\langle\psi|H^{2}|\psi\rangle-\langle\psi|H|\psi\rangle^{2}$
is the variance of the Hamiltonian $H$ with respect to the initial
state $|\psi\rangle$. If the time interval $\tau$ is so short that
$\tau\sqrt{\var}\ll1$, the effective decay rate \eqref{eq:11} becomes
\begin{equation}
\gammaeff\left(\tau\right)\approx\tau\var.\label{eq:rtau}
\end{equation}

It is interesting to observe from Eq. \eqref{eq:ptau} that when the
frequency of measurements $\nu=\tau^{-1}$ is sufficiently large,
i.e., $\tau\rightarrow0$,
\begin{equation}
\gammaeff\left(\tau\right)\rightarrow0.
\end{equation}

This is the limit of ``continuous observation'', named by Misra
and Sudarshan \citep{misra1976thezenotextquoterights}, and the survival
probability in this case turns out to be
\begin{equation}
P\left(t\right)\rightarrow1,
\end{equation}
implying that the state of the quantum system almost does not change
with time and the quantum evolution freezes. This is the quantum Zeno
effect.

Instead, if the frequency of measurements is large but still finite,
the effective decay rate will be small but finite, which means that
the final survival probability of the initial state will slowly decrease
with the evolution time $t$. Ifthe exponential decay of the survival
probability is faster than the natural decay of the quantum system
induced by noise, e.g., the amplitude damping, without repetitive
measurements, it is called quantum anti-Zeno effect.

In the past few years, it is extensively investigated how the effective
decay rate $\gammaeff\left(\tau\right)$ is influenced by the measurement
interval $\tau$ in various systems and whether it is possible to
restore the natural decay rate $\gammafree$ given by the Fermi golden
rule. The ratio of $\gammaeff\left(\tau\right)$ to $\gammafree$
is a critical factor to distinguish between the quantum Zeno effect
and the quantum anti-Zeno effect \citep{zheng2008quantum,zhang2018criterion}:
the quantum Zeno effect occurs if $\gammaeff\left(\tau\right)/\gammafree<1$,
and the quantum anti-Zeno effect occurs if $\gammaeff\left(\tau\right)/\gammafree>1$.

\section{Coherent quantum control scheme\label{sec:STRONG-QUANTUM-CONTROL}}

In this section, we consider a general quantum system with a free
Hamiltonian $\hami[0]$, suffering from Markovian noise and being
repetitively observed by a projective measurement. Starting with the
most general Markovian noise and its impact on the quantum Zeno effect,
our aim is to pursue a quantum control scheme to suppress the influence
of noise.

Generally, if the dimension of the system is large enough to prepare
a quantum error correction code for the given noise and the initial
state of the system happens to live in the code subspace, one can
use the syndrome detection and unitary recover operations of the quantum
error correction code to protect the Zeno effect. For more general
scenarios, this is not always the case, and one needs to resort to
other methods to suppress the influence of noise on the Zeno effect.
Inspired by the dynamical decoupling method, we explore coherent quantum
controls such as Hamiltonian controls to protect the Zeno effect against
noise in this paper. The dynamical decoupling requires that the control
pulses are performed sufficiently frequently so that the interval
between two consecutive control pulses is shorter than the correlation
time of the noise. The requirement for the Hamiltonian control to
protect the Zeno effect is similar here: as the magnitude of the change
of a quantum state by Markovian noise is $O(\tau)$ while the change
by a Hamiltonian is of order $O(\tau^{2})$ for a short time interval
$\tau$ between two measurements, the Hamiltonian control needs to
be as strong as of order $O(\tau^{-1})$ to suppress the influence
of the noise when the frequency of the measurements is large but finite,
i.e. $\tau$ is small but nonzero. So we will mainly consider strong
Hamiltonian controls in this section.

This section provides the necessary conditions for a coherent quantum
control scheme to be capable of suppressing the effects of Markovian
noise on the quantum Zeno effect and obtain general analytical results
for the decay rate of the survival probability in the presence of
noise with the Hamiltonian control. Moreover, we propose the ensemble
average fidelity as a metric to evaluate the overall performance of
the Hamiltonian control in protecting the quantum Zeno effect against
noise over all possible initial states of the quantum system.

It is known by the theory of open quantum systems that the evolution
of a general quantum system with a Hamiltonian and Markovian noise
can be described by the master equation
\begin{align}
\partial_{t}\rho(t)=\mathcal{L}_{\tot}\left[\rho(t)\right] & =\huaLh\left[\rho(t)\right]+\huaL[\mu]\left[\rho(t)\right],\label{eq:70}
\end{align}
where $\mathcal{L}_{\tot}$ is the total generator of the system evolution
and $\huaLh\left[\cdot\right]$, $\huaL[\mu]\left[\cdot\right]$ are
the generators of the Hamiltonian evolution and the dissipation process,
respectively,
\begin{equation}
\begin{aligned}\huaLh\left[\cdot\right]= & -i[H,\cdot],\\
\huaL[\mu]\left[\cdot\right]= & \sum_{k}\inten_{k}\huaD[V_{k}](\cdot)=\sum_{k}\inten_{k}[V_{k}(\cdot)V_{k}^{\dagger}-\frac{1}{2}\{V_{k}^{\dagger}V_{k},\cdot\}].
\end{aligned}
\label{eq:lr}
\end{equation}
The dissipation rates $\inten_{k}$'s are assumed to be non-negative
to guarantee the Markovianity of the noise \citep{rivas2012openquantum}
. For the sake of simplicity, we assume that both $H$ and $V_{k}$'s
in Eq.$\ $\eqref{eq:70} are time-independent.

The master equation \eqref{eq:70} can be formally solved by exponentiating
the total Liouvillian $\mathcal{L}_{\tot}=\huaLh+\huaL[\mu]$,
\begin{equation}
\rho\left(t\right)=e^{\huaL[tot]t}\left[\rho\left(0\right)\right],\label{eq:54}
\end{equation}
and the survival probability of the initial state of the quantum system
after an evolution of time $t$ under the Hamiltonian and the noise
between two consecutive measurements in the quantum Zeno effect is
given by
\begin{equation}
p\left(t\right)=\lef[\psio]\rho\left(t\right)\rig[\psio],\label{eq:52}
\end{equation}
where $\rig[\psio]$ is the initial state of the quantum system and
$\rho(0)$ is the density matrix of the initial state, $\rho(0)=\rig[\psio]\lef[\psio]$.

\subsection{Control-free scheme\label{subsec:Markovian-Noise-Channel-1}}

Before introducing quantum controls to reduce the impact of noise
on the quantum Zeno effect, a general quantum system with a free evolution
Hamiltonian is considered in this subsection to see the behavior of
the final survival probability of the quantum system to stay in the
initial state after repetitive projective measurements in the Zeno
limit without the protection by quantum control against the noise.

The survival probability after an evolution of $\tau$ under the master
equation \eqref{eq:70} followed by a single measurement can be written
as
\begin{equation}
\begin{aligned}p_{\noise}\left(\tau\right) & =\lef[\psio]e^{\huaL[tot]^{(\noise)}\tau}\left[\rou[0]\right]\rig[\psio]\\
 & =\lef[\psio]e^{\left(\huaLh[0]+\huaL[\inten]\right)\tau}\left[\rou[0]\right]\rig[\psio],
\end{aligned}
\end{equation}
where the subscript ``$\noise$'' denotes the absence of quantum
control to distinguish from the case with quantum control below.

When the time interval $\tau$ between two consecutive measurements
is short, the short-time behavior of $p_{\noise}\left(\tau\right)$
can be obtained by Taylor expansion to the second order of $\tau$,
\begin{align}
p_{\noise}\left(\tau\right) & =1+\lef[\psio]\huaL[\inten]\left[\rou[0]\right]\rig[\psio]\tau\nonumber \\
 & +\frac{1}{2}\lef[\psio]\left(\huaLh[0]+\huaL[\inten]\right)^{2}\left[\rou[0]\right]\rig[\psio]\tau^{2}+O\left(\tau^{3}\right).\label{eq:25}
\end{align}
There should have been another term $\lef[\psio]\huaLh[0]\left[\rou[0]\right]\rig[\psio]$
in the first-order coefficient, but it has been dropped as it is always
zero considering $\huaLh[0]$ is a commutator and $\rho_{0}=\rig[\psio]\lef[\psio]$.
When the projective measurement is performed repetitively, the final
survival probability of the system in the initial state after an evolution
time $t$ can be obtained as
\begin{equation}
P_{\noise}\left(t\right)=p_{\noise}\left(\tau\right)^{t/\tau}=\exp\left[-\gammaeff\left(\tau\right)t\right],
\end{equation}
where the subscript ``$\noise$'' in $P_{\noise}\left(t\right)$
also denotes the absence of quantum control and $\gammaeff\left(\tau\right)$
is the effective decay rate of the system,
\begin{equation}
\gammaeff\left(\tau\right)=-\frac{\ln p(\tau)}{\tau}.
\end{equation}

Substituting Eq.$\ $\eqref{eq:25} into $\gammaeff\left(\tau\right)$,
one can obtain the approximation of $\gammaeff\left(\tau\right)$
to the first order of $\tau$ as
\begin{equation}
\gammaeff^{(\noise)}\left(\tau\right)=-\langle\huaL[\inten]\rangle-\frac{1}{2}\langle\Delta^{2}\left(\huaLh[0]+\huaL[\inten]\right)\rangle\tau+O\left(\tau^{2}\right).\label{eq:26}
\end{equation}
The superscript ``$(\noise)$'' denotes the absence of quantum control,
and $\langle\huaL[\inten]\rangle$, $\langle\Delta^{2}\left(\huaLh[0]+\huaL[\inten]\right)\rangle$
denotes the mean and the variance of the superoperator $\mathcal{L}$
in the Liouville space, respectively \citep{gyamfi2020fundamentals},
with
\begin{equation}
\langle\Delta^{2}\huaL\rangle\equiv\langle\huaL^{2}\rangle-\langle\huaL\rangle^{2},
\end{equation}
and
\begin{equation}
\langle\huaL^{k}\rangle=\lef[\psio]\huaL^{k}\left[\rou[0]\right]\rig[\psio]=\tr\left(\rho_{0}\huaL^{k}\left[\rou[0]\right]\right)=\langle\overrightarrow{\rou[0]}|L^{k}|\overrightarrow{\rou[0]}\rangle,
\end{equation}
where the operator $L$ in the Roman font and the ket $|\overrightarrow{\rou[0]}\rangle$
denotes the matrix form of the superoperator $\huaL$ and the vector
form of the density operator $\rou[0]$ in the Liouville space, respectively.
Note this variance of the superoperator $\huaL$ does not always remain
non-negative as $\huaL$ is not necessarily Hermitian.

It can be further seen from Eq. \eqref{eq:26} that when the frequency
of measurements $\nu=\tau^{-1}$ is sufficiently large,
\begin{equation}
\nu\gg\abs{\frac{\langle\Delta^{2}\left(\huaLh[0]+\huaL[\inten]\right)\rangle}{2\langle\huaL[\inten]\rangle}},\label{eq:45}
\end{equation}
the linear term of $\gammaeff\left(\tau\right)$ in Eq.$\ $\eqref{eq:26}
becomes negligible and the effective decay rate becomes independent
of $\tau$. Note that the condition \eqref{eq:45} does not diverge
as $\lef[\psio]\huaL[\inten]\left[\rou[0]\right]\rig[\psio]$ is generally
nonzero. Consequently, the final survival probability after an evolution
of time $t$ can be approximated as
\begin{equation}
P_{\noise}\left(t\right)=e^{-\gammaeff^{(\noise)}t},\label{eq:22}
\end{equation}
where
\begin{equation}
\begin{aligned}\gammaeff^{(\noise)} & \approx-\lef[\psio]\huaL[\inten]\left[\rou[0]\right]\rig[\psio]\\
 & =\sum_{k}\inten_{k}\left[\lef[\psio]V_{k}^{\dagger}V_{k}\rig[\psio]-\lef[\psio]V_{k}^{\dagger}\rig[\psio]\lef[\psio]V_{k}\rig[\psio]\right].
\end{aligned}
\label{eq:reff}
\end{equation}
Note that $\gammaeff^{(\noise)}$ is always non-negative due to the
Cauchy-Schwarz inequality and the non-negativity of $\inten_{k}$'s.

An important feature of the effective decay rate $\gammaeff^{(\noise)}$
\eqref{eq:reff} is its independence of the time interval $\tau$
between two consecutive projective measurements due to the appearance
of the linear term in the expansion of $p_{\noise}\left(\tau\right)$,
which implies that the decay rate does not vanish when $\tau\rightarrow0$
and the survival probability always decays with time in this case.
This is in sharp contrast to the quantum Zeno effect where the effective
decay rate \eqref{eq:rtau} is proportional to $\tau$ and vanishes
when $\tau\rightarrow0$. It results in the failure to freeze the
evolution of the quantum system in the presence of Markovian noise,
implying the quantum Zeno effect vanishes in this case, which is consistent
with the results in the existing literature, e.g., \citep{gurvitz2003relaxation,elattari2000effectof,gurvitz2003quantum}.

\subsection{Coherent control scheme\label{subsec:Coherent-Control-Scheme}}

As Markovian noise makes the survival probability of the system to
stay in the initial state to decay exponentially even with frequent
projective measurements, it is desirable to protect the Zeno effect
against the noise with proper quantum control method. From the results
in the preceding subsection, it can be seen that the key to the exponential
decay of the survival probability lies in the the linear term of the
survival probability in the Taylor expansion introduced by the noise
after a single step of evolution and projective measurement. So the
aim of the quantum control is to decrease the linear term in the survival
probability of a single step of evolution and measurement.

\begin{figure}
\includegraphics[width=8.6cm]{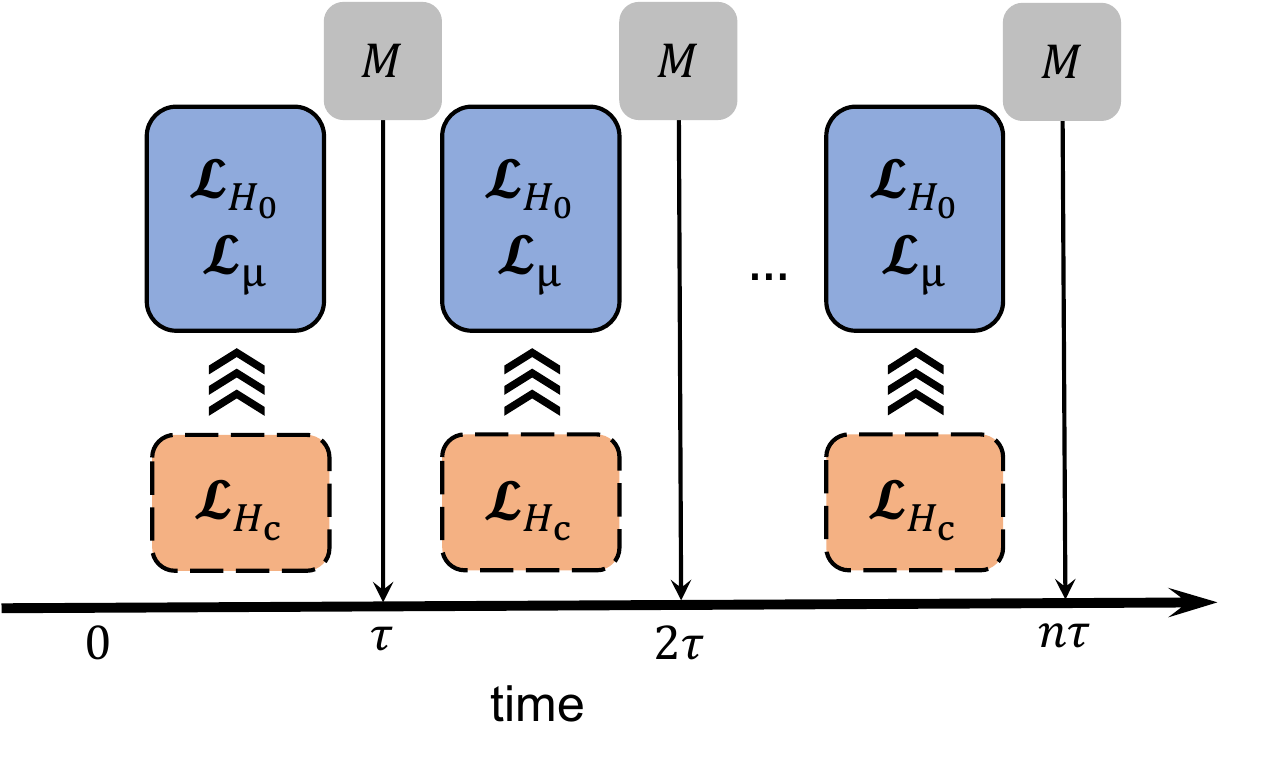}

\caption{\label{fig:0}Scheme of the coherent-control-enhanced quantum Zeno
effect in the presence of Markovian noise. In the absence of quantum
control, the quantum system undergoes a free unitary evolution and
noise simultaneously. At the end of each time interval $\tau$, a
projective measurement is performed to observe if the system remains
in its initial state. The projective measurement is assumed to be
instantaneous, implying no evolution occurs during the measurement
process. This process repeats every time interval $\tau$. In the
presence of a coherent quantum control, a proper additional control
Hamiltonian is applied on the quantum system to engineer the dynamics
of the system so that the effective decay rate of the survival probability
can be decreased. The projective measurement is denoted by $M$, and
the evolution of free Hamiltonian, the noise and the control Hamiltonian
are denoted by $\protect\huaLh[0]$,$\protect\huaL[\protect\inten]$
and $\protect\huaLh[\protect\control]$, respectively.}
\end{figure}

In this subsection, we consider a coherent quantum control scheme
to suppress the influence of noise on the quantum Zeno effect and
decrease the decay rate of the survival probability of the system
in the initial state. The evolution of the quantum system with a coherent
control scheme in the presence of noise is illustrated in Fig. \ref{fig:0}.

It should be noted that within a short but finite time interval $\tau$,
the change of the survival probability by a Hamiltonian is $O(\tau^{2})$
\eqref{eq:ptau} but the change of the survival probability by Markovian
noise is $O(\tau)$ \eqref{eq:25}, so if the purpose of a control
Hamiltonian is to suppress the effect of noise on the survival probability,
the control Hamiltonian needs to be sufficiently strong. As shown
in the following, the control Hamiltonian actually needs to be as
strong as $O(\tau^{-1})$ to slow the decay of the survival probability,
which means the total change of the system induced by the control
Hamiltonian over the short time interval $\tau$ is approximately
$O(1)$, which is similar to those fast pulse controls such as dynamical
decoupling and quantum control by reverse optimized pulse sequences
\citep{viola1998dynamical,yang2008universality,guerra2012decoherence}.

Suppose a control Hamiltonian $g\hami[\control]$ is performed on
the quantum system in the presence of noise, where $g$ is the strength
parameter. When the measurement interval is $\tau$, the short-time
survival probability of the initial state with the Hamiltonian control
after a single measurement reads
\begin{equation}
\begin{aligned}\prot[\control](\tau) & =\lef[\psio]e^{\huaL[tot]\tau}\left[\rou[0]\right]\rig[\psio]\\
 & =\lef[\psio]e^{\left(\huaLh[0]+\huaL[\inten]+g\huaLh[\control]\right)\tau}\left[\rou[0]\right]\rig[\psio],
\end{aligned}
\label{eq:29}
\end{equation}
where the subscript ``$\control$'' in $\prot[\control](\tau)$
denotes the presence of coherent quantum control.

To explore how strong the control Hamiltonian needs to be to suppress
the influence of the Markovian noise, we try the ansatz $g=\omega\tau^{k}$,
and the task is to find appropriate choices of $k$ to increase the
higher-order terms of the short-time survival probability to the first
order of $\tau$. Note that in this case, $\huaL[tot]$ will be dependent
on $\tau$, so in the following we will denote $\huaL[tot]$ as $\huaL[tot]^{(\control)}(\tau)$
to explicitly indicate its dependence on $\tau$ with the control
Hamiltonian applied. But $\huaL[tot]^{(\control)}(\tau)$ is still
independent of the instantaneous time points during the evolution
as $\tau$ is just the time interval between two consecutive projective
measurements which can be taken as a parameter, so the evolution under
$\huaL[tot]^{(\control)}(\tau)$ between two measurements can still
be written as $e^{\tau\huaL[tot]^{(\control)}(\tau)}$.

It is difficult to directly expand $\prot[\control](\tau)$ with respect
to $\tau$ by the Taylor expansion of the evolution superoperator
$e^{\tau\huaL[tot]^{(\control)}(\tau)}$ analogous to Eq. \eqref{eq:25},
\begin{equation}
e^{\tau\huaL[tot]^{(\control)}(\tau)}=\sum_{j}\frac{\left(\huaLh[0]\tau+\huaL[\inten]\tau+\omega\tau^{k+1}\huaLh[\control]\right)^{j}}{j!},
\end{equation}
since $\huaLh[0]\tau+\huaL[\inten]\tau+\omega\tau^{k+1}\huaLh[\control]$
is not proportional to $\tau$ here and all the terms of this expansion
include the lowest-order terms of $\tau$. Therefore, we turn to compute
the first few derivatives of $e^{\tau\huaL[tot]^{(\control)}(\tau)}$
with respect to $\tau$ to give the leading terms in the expansion
of $\prot[\control](\tau)$. In this case, the evolution superoperator
$e^{\tau\huaL[tot]^{(\control)}(\tau)}$ can be expanded at $\tau=0$
as
\begin{equation}
\begin{aligned}e^{\tau\huaL[tot]^{(\control)}(\tau)} & =\ddlim{e^{\tau\huaL[tot]^{(\control)}(\tau)}}+\tau\ddlim{\partial_{\tau}\left(e^{\tau\huaL[tot]^{(\control)}(\tau)}\right)}\\
 & \!\!+\frac{\tau^{2}}{2}\ddlim{\partial_{\tau}^{2}\left(e^{\tau\huaL[tot]^{(\control)}(\tau)}\right)}+O(\tau^{3}).
\end{aligned}
\label{eq:elt}
\end{equation}

Substituting $g=\omega\tau^{k}$ into the evolution superoperator
$e^{\tau\huaL[tot]^{(\control)}(\tau)}$ \eqref{eq:elt} in the Liouville
space, the first-order derivative of $e^{\mathcal{L}\tau}$ \citep{wilcox1967exponential}
can be derived as
\begin{equation}
\partial_{\tau}e^{\tau\huaL[tot]^{(\control)}(\tau)}=\intop_{0}^{1}e^{\tau\huaL[tot]^{(\control)}(\tau)\left(1-\eta\right)}\partial_{\tau}\left[\tau\huaL[tot]^{(\control)}(\tau)\right]e^{\tau\huaL[tot]^{(\control)}(\tau)\eta}d\eta.\label{eq:71}
\end{equation}
At $\tau=0$ where the evolution superoperator $e^{\tau\huaL[tot]^{(\control)}(\tau)}$
is expanded,
\begin{equation}
\begin{aligned}\ddlim{\tau\huaL[tot]^{(\control)}(\tau)}= & \ddlim{\omega\huaLh[\control]\tau^{k+1}},\\
\ddlim{\partial_{\tau}\left[\tau\huaL[tot]^{(\control)}(\tau)\right]}= & \ddlim{\left(k+1\right)\omega\huaLh[\control]\tau^{k}}+\huaL[\mu]+\huaLh[0].
\end{aligned}
\label{eq:ltexpand}
\end{equation}
It can be observed that if $k>0$, $\ddlim{\tau\huaL[tot]^{(\control)}(\tau)}=0$
and $\ddlim{\partial_{\tau}\left[\tau\huaL[tot]^{(\control)}(\tau)\right]}=\huaL[\mu]+\huaLh[0]$,
so Eq. \eqref{eq:71} can be simplified to
\begin{equation}
\ddlim{\partial_{\tau}\left(e^{\tau\huaL[tot]^{(\control)}(\tau)}\right)}=\ddlim{\partial_{\tau}\left[\tau\huaL[tot]^{(\control)}(\tau)\right]}=\huaL[\mu]+\huaLh[0].
\end{equation}

In this case, the first-order terms of $e^{\tau\huaL[tot]^{(\control)}(\tau)}$
and of the short-time survival probability $\prot[\control](\tau)$
are independent of the control Hamiltonian $\hami[\control]$, so
$\hami[\control]$ cannot help decrease the decay rate of the survival
probability in the long-term evolution.

If $-1<k\leq0$, $\ddlim{\tau\huaL[tot]^{(\control)}(\tau)}$ is still
zero, but $\ddlim{\partial_{\tau}\left[\tau\huaL[tot]^{(\control)}(\tau)\right]}$
includes the control Hamiltonian now. It seems that the Hamiltonian
control is possible to decrease the decay rate of the survival probability
in this case. Nevertheless, it can be verified that for any arbitrary
initial state $\rig[\psio]$,
\begin{equation}
\lef[\psio]\huaLh[\control]\left[\rou[0]\right]\rig[\psio]=0,
\end{equation}
so the decay rate can still not be lowered by the Hamiltonian control
in this case.

If $k<-1$, it can be immediately inferred from Eq. \eqref{eq:ltexpand}
that both $\tau\huaL[tot]^{(\control)}(\tau)$ and $\partial_{\tau}\left[\tau\huaL[tot]^{(\control)}(\tau)\right]$
diverge at $\tau=0$. So, the only possible choice of $k$ to make
the Hamiltonian control scheme to work is $k=-1$, i.e., the strength
parameter of the control Hamiltonian is
\begin{equation}
g=\omega\tau^{-1}.
\end{equation}
In this case, 
\begin{equation}
\begin{aligned}\ddlim{\tau\huaL[tot]^{(\control)}(\tau)} & =\omega\huaLh[\control],\\
\ddlim{\partial_{\tau}\left[\tau\huaL[tot]^{(\control)}(\tau)\right]} & =\huaL[\mu]+\huaLh[0],
\end{aligned}
\label{eq:whole-3}
\end{equation}
where $\ddlim{\tau\huaL[tot]^{(\control)}(\tau)}$ includes the control
Hamiltonian now, so it is possible to modulate the first-order derivative
of $e^{\tau\huaL[tot]^{(\control)}(\tau)}$ by the coherent control
scheme. Hence, within the framework of coherent control schemes, the
strength of the control Hamiltonian needs to be proportional to the
frequency of the repetitive projective measurements. This means that
the total change of the system made by the control Hamiltonian is
of order $O(1)$, which is analogous to the pulse controls employed
in other quantum control tasks. And while the measurement frequency
is large in the Zeno effect, it is still finite in practice, so it
provides the feasibility of implementing this coherent control scheme
in experiments.

Substituting the strength of coherent control $g=\omega\tau^{-1}$
into the Liouvillian superoperator $\huaL[tot]^{(\control)}$, the
total evolution superoperator in the coherent control scheme can be
simplified as
\begin{equation}
e^{\tau\huaL[tot]^{(\control)}(\tau)}=e^{\omega\huaLh[\control]+(\huaL[\mu]+\huaLh[0])\tau}.
\end{equation}
The survival probability after a single measurement is can be expanded
at $\tau=0$ as
\begin{align}
\begin{aligned}\prot[\control](\tau) & =\prot[\control](0)+\partial_{\tau}\prot[\control](\tau)|_{\tau=0}\tau\\
 & \ \ +\partial_{\tau}^{2}\prot[\control](\tau)|_{\tau=0}\frac{\tau^{2}}{2}+O(\tau^{3}),
\end{aligned}
\label{eq:59}
\end{align}
where $\pt[k]\prot[\control](\tau)$ is the $k$th derivative of survival
probability $\prot[\control](\tau)$ at $\tau=0$,
\begin{equation}
\pt[k]\prot[\control](\tau)=\lef[\psio](\pt[k]e^{\tau\huaL[tot]^{(\control)}(\tau)})\left[\rou[0]\right]\rig[\psio].\label{eq:ptpct}
\end{equation}

We first compute the zeroth-order term in the Taylor expansion of
$\prot[\control](\tau)$ \eqref{eq:59},
\begin{equation}
\prot[\control]|_{\tau=0}=\lef[\psio]e^{\omega\huaLh[\control]}\left[\rou[0]\right]\rig[\psio].
\end{equation}
By the definition of $\huaLh[\control]$, $\huaLh[\control]\left[\cdot\right]=-i[\hami[\control],\cdot]$,
$\prot[\control]|_{\tau=0}$ can be written as
\begin{equation}
\prot[\control]|_{\tau=0}=|\lef[\psio]e^{-i\omega\hami[\control]}\rig[\psio]|^{2}.
\end{equation}

To ensure that the zeroth-order term of $\prot[\control]|_{\tau=0}$
remains 1 for an arbitrary initial state $\rig[\psio]$, $e^{-i\omega\hami[\control]}$
needs to satisfy
\begin{equation}
e^{-i\omega\hami[\control]}=e^{i\theta}I,
\end{equation}
where $e^{i\theta}$ is an arbitrary phase and $I$ is the identity
operator. This requires that
\begin{equation}
\omega\big(E_{i}^{(\control)}-E_{j}^{(\control)}\big)=2n\pi,\;n=\pm1,\pm2,\cdots,\;\forall i\neq j,\label{eq:16}
\end{equation}
where $E_{i}^{(\control)}$ and $E_{j}^{(\control)}$ are two arbitrary
eigenvalues of the control Hamiltonian $\hami[c]$, which immediately
leads to a necessary condition for the control Hamiltonian to preserve
the initial state in the limit $\tau\rightarrow0$,
\begin{equation}
\Delta E_{ij}^{(\control)}/\Delta E_{i^{\prime}j^{\prime}}^{(\control)}\in\mathbb{Q},\;\forall i\neq j,i^{\prime}\neq j^{\prime},
\end{equation}
where $\Delta E_{ij}^{(\control)}=E_{i}^{(\control)}-E_{j}^{(\control)}$
and $\mathbb{Q}$ is the set of all rational numbers. When this condition
is satisfied, $e^{\omega\huaLh[\control]}$ can be simplified to
\begin{equation}
e^{\omega\huaLh[\control]}=\mathcal{I},\label{eq:superid}
\end{equation}
where $\mathcal{I}$ is the identity superoperator in the Liouville
space.

Back to the derivatives of the evolution superoperator $e^{\huaL[tot]^{(\control)}\tau}$,
by substituting Eqs. \eqref{eq:whole-3} and \eqref{eq:superid} into
Eq.$\ $\eqref{eq:71}, the first-order derivative of $e^{\tau\huaL[tot]^{(\control)}(\tau)}$
at $\tau=0$ can be rewritten as
\begin{equation}
\ddlim{\partial_{\tau}e^{\tau\huaL[tot]^{(\control)}(\tau)}}=\int_{0}^{1}e^{-\omega\huaLh[\control]\eta}\left(\huaL[\mu]+\huaLh[0]\right)e^{\omega\huaLh[\control]\eta}d\eta,\label{eq:72}
\end{equation}
and the second-order derivative of $e^{\tau\huaL[tot]^{(\control)}(\tau)}$
can also be obtained,
\begin{equation}
\begin{aligned} & \ddlim{\partial_{\tau}^{2}e^{\tau\huaL[tot]^{(\control)}(\tau)}}\\
 & =\int_{0}^{1}d\eta_{2}\int_{0}^{\eta_{2}}e^{-\omega\huaLh[\control]\eta_{2}}\left(\huaL[\mu]+\huaLh[0]\right)e^{\omega\huaLh[\control]\left(\eta_{2}-\eta_{1}\right)}\\
 & \ \ \times\left(\huaL[\mu]+\huaLh[0]\right)e^{\omega\huaLh[\control]\eta_{1}}d\eta_{1},
\end{aligned}
\end{equation}
which will be useful in deriving the condition for the frequency of
the measurements below.

It can be straightforwardly verified that
\begin{equation}
\begin{aligned}e^{-\omega\huaLh[\control]\eta}\huaLh[0]e^{\omega\huaLh[\control]\eta} & =\widetilde{\mathcal{L}}_{H_{\mathrm{0}}}^{(\eta)}=\mathcal{L}_{\widetilde{H}_{\mathrm{0}}\left(\eta\right)},\\
e^{-\omega\huaLh[\control]\eta}\huaL[\inten]e^{\omega\huaLh[\control]\eta} & =\widetilde{\mathcal{L}}_{\zheng{\inten}}^{(\eta)}=\sum_{k}\inten_{k}\huaD[\widetilde{V_{k}}(\eta)],
\end{aligned}
\label{eq:ele}
\end{equation}
where
\begin{equation}
\begin{aligned}\widetilde{H}_{0}(\eta)= & e^{i\omega\hami[\control]\eta}\hami[0]e^{-i\omega\hami[\control]\eta},\\
\widetilde{V_{k}}(\eta)= & e^{i\omega\hami[\control]\eta}V_{k}e^{-i\omega\hami[\control]\eta}.
\end{aligned}
\label{eq:hve}
\end{equation}
The derivation of the above representation transformations of the
superoperators $\huaLh[0]$ and $\huaL[\inten]$ in the Liouville
space is presented in Appendix \eqref{sec:Derivation-of-the-1}. So
the first-order and second-order derivatives of $e^{\huaL[\control]\tau}$
can be simplified as
\begin{align}
\ddlim{\partial_{\tau}e^{\tau\huaL[tot]^{(\control)}(\tau)}}= & \int_{0}^{1}\left(\widetilde{\mathcal{L}}_{H_{\mathrm{0}}}^{(\eta)}+\widetilde{\mathcal{L}}_{\zheng{\inten}}^{\zheng{(\eta)}}\right)d\eta,\label{eq:1std}\\
\ddlim{\partial_{\tau}^{2}e^{\tau\huaL[tot]^{(\control)}(\tau)}}= & \int_{0}^{1}d\eta_{2}\int_{0}^{\eta_{2}}\left(\widetilde{\mathcal{L}}_{H_{\mathrm{0}}}^{(\eta_{2})}+\widetilde{\mathcal{L}}_{\zheng{\inten}}^{\zheng{(\eta_{2})}}\right)\nonumber \\
 & \times\left(\widetilde{\mathcal{L}}_{H_{\mathrm{0}}}^{(\eta_{1})}+\widetilde{\mathcal{L}}_{\zheng{\inten}}^{\zheng{(\eta_{1})}}\right)d\eta_{1}.\label{eq:2ndd}
\end{align}

The first-order derivative of survival probability \eqref{eq:ptpct}
can then be derived by substituting Eq. \eqref{eq:1std} into \eqref{eq:ptpct}
with $k=1$,
\begin{equation}
\pt\prot[\control](\tau)|_{\tau=0}=\int_{0}^{1}\lef[\psio]\widetilde{\mathcal{L}}_{\zheng{\inten}}^{\zheng{(\eta)}}\left[\rou[0]\right]\rig[\psio]d\eta,\label{eq:61}
\end{equation}
where the fact that the average of a commutator over any quantum state
is zero, i.e., $\lef[\psio]\mathcal{L}_{\widetilde{H}_{\mathrm{0}}\left(\eta\right)}\left[\rho_{0}\right]\rig[\psio]=0$,
has been considered. Hence, the effective decay rate of survival probability
$\gammaeff\left(\tau\right)$ can be found by substituting Eq. \eqref{eq:59}
into \eqref{eq:11} as 
\begin{equation}
\begin{aligned}\gammaeff^{(\control)}\left(\tau\right) & =-\pt\prot[\control](\tau)|_{\tau=0}\\
 & -\frac{1}{2}\dlim{\left[\pt[2]\prot[\control](\tau)-\left(\pt\prot[\control](\tau)\right)^{2}\right]}\tau+O\left(\tau^{2}\right).
\end{aligned}
\end{equation}

Similarly, the second-order derivative of $\prot[\control](\tau)$
can be derived by Eq. \eqref{eq:2ndd} as
\begin{equation}
\begin{aligned}\pt[2]\prot[\control](\tau)|_{\tau=0} & =\int_{0}^{1}d\eta_{2}\int_{0}^{\eta_{2}}\lef[\psio]\left(\widetilde{\mathcal{L}}_{H_{\mathrm{0}}}^{(\eta_{2})}+\widetilde{\mathcal{L}}_{\zheng{\inten}}^{\zheng{(\eta_{2})}}\right)\\
 & \ \ \times\left(\widetilde{\mathcal{L}}_{H_{\mathrm{0}}}^{(\eta_{1})}+\widetilde{\mathcal{L}}_{\zheng{\inten}}^{\zheng{(\eta_{1})}}\right)\left[\rou[0]\right]\rig[\psio]d\eta_{1}.
\end{aligned}
\end{equation}
If the frequency of measurements $\nu=\tau^{-1}$ is required to be
sufficiently large to drop the second and higher-order terms in $\gammaeff^{(\control)}\left(\tau\right)$,
i.e., to reach the Zeno limit, the frequency of the projective measurements
needs to satisfy
\begin{equation}
\nu\gg\abs{\frac{\pt[2]\prot[\control](\tau)-\left(\pt\prot[\control](\tau)\right)^{2}}{2\pt\prot[\control](\tau)}}_{\tau=0}.\label{eq:62}
\end{equation}
In this case, the survival probability in the long-term evolution
of time $t$ can be written as
\begin{align}
P_{\zheng{\control}}(t) & \approx e^{-\gammaeff^{(\control)}t},\label{eq:51}
\end{align}
where $\gammaeff^{(\control)}$ is the simplified effective decay
rate on the condition \eqref{eq:62},
\begin{equation}
\gammaeff^{(\control)}=-\pt\prot[\control](\tau)|_{\tau=0}=-\int_{0}^{1}\lef[\psio]\widetilde{\mathcal{L}}_{\zheng{\inten}}^{\zheng{(\eta)}}\left[\rou[0]\right]\rig[\psio]d\eta,\label{eq:refc}
\end{equation}
with $\widetilde{\mathcal{L}}_{\zheng{\inten}}^{\zheng{(\eta)}}$
given by Eq. \eqref{eq:ele}.

Obviously, for a given initial state $|\psi_{0}\rangle$, different
control Hamiltonians will lead to different effective decay rates
$\gammaeff^{(\control)}$, so it is desirable to optimize the control
Hamiltonian to reach the minimum effective decay rate. To benchmark
the performance of the coherent control scheme on protecting the Zeno
effect, we define the following ratio to quantify the extent to which
the control scheme can decrease the effective decay rate of the survival
probability in the presence of noise:
\begin{align}
\kappa & \equiv\frac{\gammaeff^{(\control)}}{\gammaeff^{(\noise)}}.\label{eq:67}
\end{align}
When the control slows down the decay of the quantum state, $\kappa$
is smaller than $1$, and vice versa. And the smaller the ratio $\kappa$
is, the better the coherent control scheme works.

Moreover, for given noise, the ensemble average fidelity $F$, which
is the average fidelity between the initial state of the system and
the final state evolved by the noisy quantum process over all possible
initial states \citep{nielsen2000quantum}, can be invoked to characterize
the overall performance of the coherent control scheme in lowering
the effective decay rate for all possible initial states $|\psi_{0}\rangle$
of the quantum system with a probability distribution $q(|\psi_{0}\rangle)$,
\begin{equation}
F\left(t\right)=\int q(|\psi_{0}\rangle)P_{|\psi_{0}\rangle}\left(t\right)d|\psio\rangle.\label{eq:ensemblefidlity}
\end{equation}

The ensemble average fidelity $F(t)$ decreases with time $t$ as
the survival probability $P_{|\psi_{0}\rangle}\left(t\right)$ of
each initial state $|\psi_{0}\rangle$ decreases exponentially with
time, and the slower $F(t)$ decays with time, the better the protection
effect is.

The ratio $\kappa$ and the ensemble average fidelity $F(t)$ defined
above will be employed in the next section to quantify the performance
of the coherent control scheme for a two-level system with typical
Markovian noise.

\section{Coherent quantum control scheme for two-level systems\label{sec:STRONG-QUANTUM-CONTROL-1}}

To illustrate how the influence of noise on the Zeno effect can be
suppressed by coherent quantum controls, we consider a two-level quantum
system as an example in this section.

We derive the detailed necessary condition for a coherent control
scheme to preserve the Zeno effect of a two-level system in the presence
of Markovian noise and obtain the effective decay rates of the survival
probability with frequent repetitive projective measurements in both
control-free and controlled scenarios. For the typical Markovian noise
dephasing and amplitude damping, we find the optimal control Hamiltonians
analytically and show the improvement of slowing the decay of survival
probability by the coherent control scheme. Additionally, we investigate
the performance of the coherent control scheme for initial states,
and show the relation between the improvement of survival probability
and the initial state by numerical illustrations.

\subsection{Control-free scheme\label{subsec:Markovian-Noise-Channel}}

First of all, we denote the excited state and the ground state of
the two-level system, which is subject to free Hamiltonian evolution
along with dissipative process, as $|1\rangle$ and $|0\rangle$,
respectively. The density matrix of a two-level quantum system can
generally be written as
\begin{equation}
\rho=\left(I+\boldsymbol{r\cdot\sigma}\right)/2,
\end{equation}
where $\boldsymbol{r}=\left(r_{x},r_{y},r_{z}\right)$ is called Bloch
vector with $\abs{\boldsymbol{r}}\leq1$ and $\vp=\left(\sigma_{x},\sigma_{y},\sigma_{z}\right)$
is the collection of the three Pauli operators as a vector. The excited
and ground states $|1\rangle$ and $|0\rangle$ are the eigenstates
of $\sigma_{z}$ with eigenvalues $-1$ and $1$, respectively.

In the Zeno effect, we start with a pure state $|\psi_{0}\rangle$
for the system, the density matrix of which can be denoted as
\begin{equation}
\rou[0]=|\psi_{0}\rangle\langle\psi_{0}|=\frac{I+\boldsymbol{r_{0}\cdot\sigma}}{2},
\end{equation}
where $\boldsymbol{r_{0}}$ must be a unit vector and can be written
as
\begin{equation}
\boldsymbol{r_{0}}=\left(\sin\alpha\cos\beta,\sin\alpha\sin\beta,\cos\alpha\right).\label{eq:r0}
\end{equation}

The specific form of the dissipative term $\huaL[\mu]\left[\rho\right]$
in the master equation \eqref{eq:70} for general Markovian noise
on a single qubit can be written as

\begin{equation}
\huaL[\mu]\left[\rho\right]=\sum_{ij}\inten_{ij}\left(\sigma_{i}\rho\sigma_{j}-\frac{1}{2}\left\{ \sigma_{j}\sigma_{i},\rho\right\} \right),\;i,j=x,y,z,\label{eq:47}
\end{equation}
where the coefficient matrix consisting of $\inten_{ij}$ as its elements,
\begin{equation}
(\Gamma)_{ij}=\inten_{ij},\label{eq:gammatrix}
\end{equation}
needs to be positive semidefinite in order to guarantee the Markovianity
of the noise.

According to Sec.$\ $\ref{subsec:Markovian-Noise-Channel-1}, the
effective decay rate of the final survival probability after an evolution
of time $t$ with noise \eqref{eq:47} and frequent projective measurements
in the Zeno limit could obtain as
\begin{equation}
\begin{aligned}\gammaeff^{(\noise)} & =-\lef[\psio]\huaL[\inten]\left[\rho_{0}\right]\rig[\psio]\\
 & =-\boldsymbol{r_{0}^{\zhuan}}\Gamma\boldsymbol{r_{0}}+\tr\Gamma+\nb\cdot\boldsymbol{r_{0}},
\end{aligned}
\label{eq:18}
\end{equation}
\textbf{$\nb$} is a vector determined by the imaginary parts of the
off-diagonal elements of the dissipation coefficient matrix $\Gamma$,
\begin{equation}
\nb=2(\im\mu_{23},\,\im\mu_{31},\,\im\mu_{12}),\label{eq:g}
\end{equation}
where $\im$ denotes the imaginary part of a complex number. The derivation
of Eq.$\ $\eqref{eq:18} is given in Appendix$\ $\ref{sec:Derivation-of-effective}.

\subsection{Coherent control scheme\label{subsec:Scheme-Introduced-Channel}}

In this subsection, we consider a two-level system undergoing a general
Hamiltonian $\hami[0]$ and general Markovian noise described by the
dissipative term \eqref{eq:47}, and apply a control Hamiltonian $g\hami[\control]$
with strength $g=\omega\tau^{-1}$ to suppress the influence of the
noise, where $\hami[\control]=\boldsymbol{n_{\control}}\cdot\boldsymbol{\sigma}$
and $\boldsymbol{n_{\control}}$ is a unit vector denoted as
\begin{equation}
\boldsymbol{n_{\control}}=(\sin\thetakong[\control]\cos\ph[\control],\sin\thetakong[\control]\sin\ph[\control],\cos\thetakong[\control]),
\end{equation}
which is the direction of the control Hamiltonian in the Bloch representation.

According to Sec.$\ $\ref{sec:STRONG-QUANTUM-CONTROL}, a necessary
condition \eqref{eq:16} is imposed on the control Hamiltonian $g\hami[\control]$
to ensure the zeroth order of the survival probability $p(\tau)$
remaining 1. Specifically for a two-level quantum system, this condition
turns to be
\begin{equation}
\omega=n\pi,\;n=\pm1,\pm2,\cdots\ .\label{eq:69}
\end{equation}

With the specific form of the dissipative superoperator $\huaL[\mu]$
for a two-level system given in Eq.$\ $\eqref{eq:47}, the effective
decay rate of the survival probability after a single projective measurement
with a coherent quantum control can be obtained from Eqs. \eqref{eq:61}
and \eqref{eq:refc} as
\begin{align}
\gammaeff^{(\control)}= & -\pt\prot[\control](\tau)|_{\tau=0},\nonumber \\
= & -\sum_{i,j=1}^{3}\inten_{ij}\int_{0}^{1}\tr\left[\rou[0]\widetilde{\sigma}_{i}^{(\eta)}\rou[0]\widetilde{\sigma}_{j}^{(\eta)}-\rou[0]\widetilde{\sigma}_{j}^{(\eta)}\widetilde{\sigma}_{i}^{(\eta)}\right]d\eta,\label{eq:68}
\end{align}
where $\widetilde{\sigma}_{i}^{(\eta)}=e^{i\omega\hami[\control]\eta}\sigma_{i}e^{-i\omega\hami[\control]\eta}$
denotes the $i$th Pauli operator in the framework rotated by $e^{-i\omega\hami[\control]\eta}$
dependent on the parameter $\eta$.

When the necessary condition \eqref{eq:69} is met and the measurement
frequency reaches the Zeno limit, the effective decay rate of the
survival probability can be worked out by Eq. \eqref{eq:68} as
\begin{equation}
\begin{aligned}\gammaeff^{(\control)} & =-\frac{3}{2}(\boldsymbol{n_{\control}}\cdot\boldsymbol{r_{0}})^{2}\boldsymbol{n_{\control}^{\zhuan}}\Gamma\boldsymbol{n_{\control}}+\frac{1}{2}(\boldsymbol{n_{\control}}\cdot\boldsymbol{r_{0}})(\boldsymbol{n_{\control}^{\zhuan}}\Gamma\boldsymbol{r_{0}}+\boldsymbol{r_{0}^{\zhuan}}\Gamma\boldsymbol{n_{\control}})\\
 & \ \ -\frac{1}{2}\boldsymbol{r_{0}^{\zhuan}}\Gamma\boldsymbol{r_{0}}-\frac{1}{2}\left(\boldsymbol{n_{\control}}\times\boldsymbol{r_{0}}\right)^{\zhuan}\Gamma\left(\boldsymbol{n_{\control}}\times\boldsymbol{r_{0}}\right)\\
 & \ \ +\tr\Gamma+(\boldsymbol{n_{\control}}\cdot\boldsymbol{r_{0}})(\nb\cdot\boldsymbol{n_{\control}}),
\end{aligned}
\label{eq:66}
\end{equation}
where \textbf{$\nb$} is the vector defined in Eq. \eqref{eq:g} and
the superscript ``$\zhuan$'' denotes the matrix transposition.
The detail of derivation is provided in Appendix$\ $\ref{sec:Derivation-of-effective}.

It can be observed that when $\boldsymbol{n_{\control}}\cdot\boldsymbol{r_{0}}=\pm1$,
i.e., the direction of the Hamiltonian $\boldsymbol{n_{\control}}$
is parallel or anti-parallel with that of the initial state $\boldsymbol{r_{0}}$
since both $\boldsymbol{n_{\control}}$ and $\boldsymbol{r_{0}}$
are unit vectors, the effective decay rate $\gammaeff^{(\control)}$
in Eq.$\ $\eqref{eq:66} will coincide with $\gammaeff^{(\noise)}$
without quantum control in Eq.$\ $\eqref{eq:18}, which means the
coherent control does not have any effect on the decay rate in this
case, leading to another necessary condition for the validity of the
coherent control scheme on a two-level system,
\begin{equation}
\boldsymbol{n_{\control}}\neq\pm\boldsymbol{r_{0}}.\label{eq:76}
\end{equation}

By evaluating the ratio $\kappa$ defined in Eq.$\ $\eqref{eq:67}
with the results for $\gammaeff^{(\noise)}$ and $\gammaeff^{(\control)}$
in Eqs.$\ $\eqref{eq:18} and \eqref{eq:66}, one can determine the
effect of the coherent control scheme on the two-level system. In
particular, if $\kappa>1$, the decay of the survival probability
accelerates, and if $\kappa<1$, the decay slows down.

It can be observed from Eq. \eqref{eq:66}, that for given noise $\huaL[\inten]$'s
and an initial state $\rou[0]$, the effective decay rate $\gammaeff^{(\control)}$
of a two-level system varies with different quantum control Hamiltonian
$\hami[\control]$, so one can optimize the control Hamiltonian to
minimize the decay rate of the survival probability. The optimization
for the control Hamiltonian $\hami[\control]$ can be formally solved
by a variational approach, with the Lagrangian function constrained
by the normalization condition $\boldsymbol{n_{\control}}=1$ as
\begin{equation}
L\left(P_{\zheng n},\Lambda\right)=\gammaeff^{(\control)}+\tr\left[\left(P_{\zheng n}^{2}-P_{\zheng n}\right)\Lambda\right],\label{eq:Lagr}
\end{equation}
where the Lagrange multiplier $\Lambda$ is an arbitrary matrix and
$P_{\zheng n}\equiv\boldsymbol{n_{\control}}\boldsymbol{n_{\control}^{\zhuan}}$
is the projection operators on to the unit vector $\boldsymbol{n_{\control}}$
in the three-dimensional Bloch space.

To obtain the optimal control Hamiltonian to minimize the effective
decay rate $\gammaeff^{(\control)}$ of the two-level quantum system,
the variation of the Lagrangian function \eqref{eq:Lagr} needs always
to be zero for any $\delta P_{\zheng n}$ and $\delta\Lambda$ according
to the principle of variation approach, leading to the optimization
equation for the projector $P_{\zheng n}$ onto the unit vector $\boldsymbol{n_{\control}}$
along the direction of the control Hamiltonian,

\begin{align}
-\frac{3}{2}\Gamma P_{\zheng n}P_{\zheng r}-\frac{3}{2}P_{\zheng r}P_{\zheng n}\Gamma+\frac{1}{2}\Gamma P_{\zheng r}+\frac{1}{2}P_{\zheng r}\Gamma\nonumber \\
-\frac{1}{2}R^{\zhuan}\Gamma R+\boldsymbol{r_{0}}\nb^{\zhuan}+P_{\zheng n}\Lambda+\Lambda P_{\zheng n}-\Lambda & =\mathbf{0},\label{eq:81}
\end{align}
with the constraint condition $P_{\zheng n}^{2}=P_{\zheng n}$, where
$P_{\zheng r}=\boldsymbol{r_{0}}\boldsymbol{r_{0}^{\zhuan}}$ is the
projection operators onto the vector $\boldsymbol{r_{0}}$ in the
Bloch space and $R$ is an antisymmetric matrix,
\begin{equation}
R=\begin{bmatrix}0 & z_{0} & -y_{0}\\
-z_{0} & 0 & x_{0}\\
y_{0} & -x_{0} & 0
\end{bmatrix},
\end{equation}
with $\boldsymbol{r_{0}}=(x_{0},y_{0},z_{0})$. The detail of derivation
is provided in Appendix \ref{subsec:two-level-quantum-system}. Once
the optimization equation \eqref{eq:81} is solved, the direction
of the optimal control Hamiltonian can be determined, and the optimal
control Hamiltonian can be obtained by this direction with the leading
factor given in Eq. \eqref{eq:69}.

\subsection{Examples\label{subsec:Application}}

To illustrate the above general theoretical results, we investigate
the effect of coherent quantum controls on two-level quantum systems
with two typical types of Markovian noise, the dephasing and the amplitude
damping. By deriving the effective decay coefficients $\gammaeff^{(\noise)}$
and $\gammaeff^{(\control)}$ under these specific noise channels,
we find the optimal coherent controls tailored for different initial
states of two-level quantum systems. Furthermore, we devote to unveiling
the physical pictures underlying the optimal coherent control strategy
for each type of noise. Through a comparative analysis of the decay
rate and the ensemble average fidelity between the cases with or without
the optimal coherent control in the presence of noise, we demonstrate
the advantage of this coherent control scheme in protecting the Zeno
effect both analytically and numerically.

In addition, we will also consider the impact of different initial
states of the quantum system, the improvement of the effective decay
rates brought by the coherent quantum controls, and study the relation
between the initial state and the extent to which the decay rate can
be decreased by coherent control in detail through numerical computation.

\subsubsection{Dephasing}

The dephasing noise is a typical unitary noise described by a dissipative
term $\huaD[\sigma_{z}]$ in a quantum master equation, and has been
extensively studied in the theory of open quantum systems, with the
noise coefficient matrix \eqref{eq:gammatrix} as
\begin{equation}
\Gamma_{\zheng z}=\left(\begin{array}{ccc}
0 & 0 & 0\\
0 & 0 & 0\\
0 & 0 & \inten
\end{array}\right).\label{eq:24}
\end{equation}
The evolution of a two-level quantum system under a free Hamiltonian
$\hami[0]$ and the dephasing noise with an intensity $\mu$ is determined
by the master equation
\begin{equation}
\frac{d\rot}{dt}=-i\left[\hami[0],\rot\right]+\inten\huaD[\sigma_{z}]\rot,\label{eq:31}
\end{equation}
where $\huaD[\sigma_{z}]\rot=\sigma_{z}\rot\sigma_{z}-\rot$ and $\rot$
is the density matrix of the system at time $t$. When a control Hamiltonian
$g\hami[\control]$ is introduced to the system, the master equation
becomes
\begin{equation}
\frac{d\rot}{dt}=-i\left[\hami[0]+g\hami[\control],\rot\right]+\inten\huaD[\sigma_{z}]\rot.\label{eq:meq}
\end{equation}

\begin{figure}
\includegraphics[width=4.3cm]{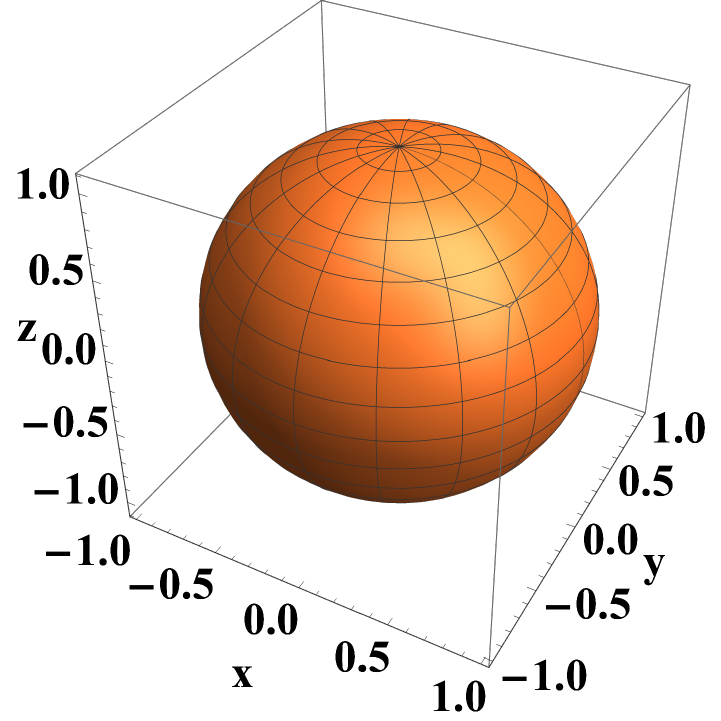}\includegraphics[width=4.3cm]{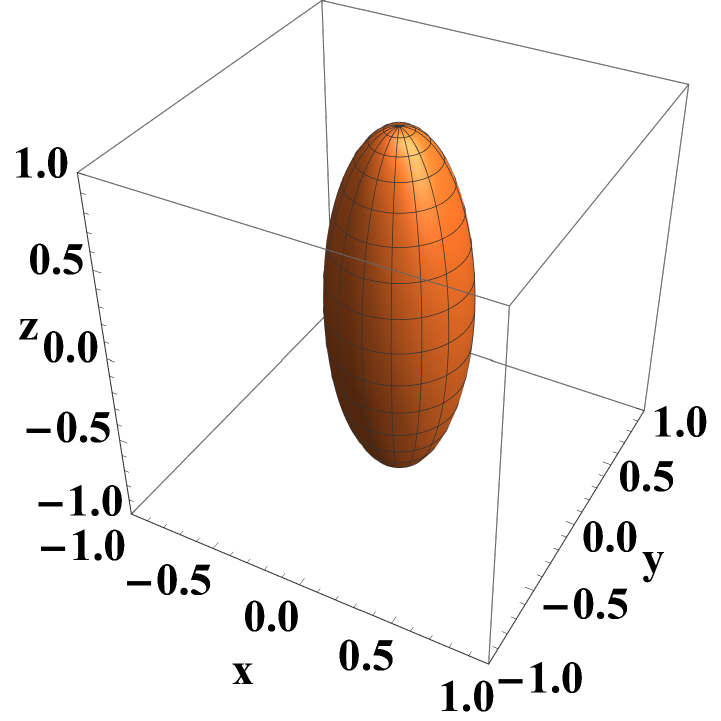}

\caption{\label{fig:1}Transformation of the Bloch sphere under the dephasing
noise. The dephasing noise results in the invariance of $z$ components
of the Bloch vectors and uniform contraction of the $x$ and $y$
towards the $z$-axis. Parameter: $\protect\inten t=1/2$.}
\end{figure}

The effect of the dephasing noise on a two-level system is plotted
in Fig. \ref{fig:1}. It shows that the Bloch sphere, which includes
all possible density matrices of a two-level system, is ``compressed''
towards the $z$-axis by the dephasing noise. And the compression
is symmetric about the equatorial plane as the dephasing noise does
not change the $\sigma_{z}$ component of any density matrix and rotationally
symmetric around the $z$-axis as the dephasing noise affects the
$\sigma_{x}$ and $\sigma_{y}$ components of all density matrices
uniformly.

When the frequence of the projetive measurement is large enough and
the condition \eqref{eq:69} is satisfied in the Zeno limit $\tau\rightarrow0$,
by substituting the $\boldsymbol{r_{0}}$ \eqref{eq:r0} and $\Gamma_{\zheng z}$
\eqref{eq:24} into the Eqs.$\ $\eqref{eq:18} and \eqref{eq:66},
one can obtain the effective dacay rates of survival probability without
the coherent control,
\begin{equation}
\gammaeff^{(\noise)}=\frac{\inten}{2}\left(1-\cos2\alpha\right),\label{eq:32}
\end{equation}
and with the coherent control
\begin{equation}
\begin{aligned}\gammaeff^{(\control)}= & \frac{\inten}{64}\Big[39-2\cos2\alpha\left(1+3\cos2\thetakong[\control]\right)^{2}\\
 & -3\cos4\thetakong[\control]-8\cos2\Delta\sin^{2}\alpha\sin^{2}\thetakong[\control]\\
 & -4\cos2\thetakong[\control]\left(1+6\cos2\Delta\sin^{2}\alpha\sin^{2}\thetakong[\control]\right)\\
 & -4\cos\Delta\sin2\alpha\left(2\sin2\thetakong[\control]+3\sin4\thetakong[\control]\right)\Big],
\end{aligned}
\label{eq:42}
\end{equation}
where $\Delta=\beta-\ph[\control]$.

From the result of $\gammaeff^{(\noise)}$ \eqref{eq:32}, it is evident
that the decay rate of survival probability induced by the dephasing
noise is independent of the azimuthal angle $\beta$ and solely dependent
on the polar angle $\alpha$ between the initial state and the $z$-axis,
and is symmetric about $\alpha=\pi/2$, in agreement with the symmetries
of the impact of the dephasing noise on the Bloch sphere shown in
Fig.$\ $\ref{fig:1}. When the angle $\alpha$ is zero or $\pi$,
the initial state remains unaffected by the noise as it lies along
the compression axis, i.e., the $z$ axis, so no decay occurs in the
survival probability in this case. However, when the polar angle $\alpha$
is $\pi/2$, the effect of noise becomes maximal, and the effective
decay rate increases to $\inten/2$. An intriguing discovery emerges
from Eq. \eqref{eq:42}: the ratio $\kappa=\gammaeff^{(\control)}/\gammaeff^{(\noise)}$
with the coherent control optimized is independent of the noise intensity
$\inten$, suggesting the coherent control scheme is robust against
different strengths of the dephasing noise, which is a desirable property
for application of this control scheme in real environments.

Due to the different impacts of the dephasing noise on different initial
states of the two-level system, the extent to which the decay of the
survival probability can be decreased by coherent quantum controls
is also different. Obviously, when the Bloch vector of the initial
state lies along the $z$ axis, i.e., $\alpha=0,\;\pi$, the dephasing
noise does not change the initial state, so any coherent quantum control
cannot improve the probability for the system to stay in the initial
state if it does not even worsen the decay of survival probability.
Note that the free Hamiltonian may rotate the initial state from the
$z$-axis to another direction that suffers from the dephasing noise,
but as repetitive projective measurements are performed on the system
with a sufficiently large frequency, the change of the system induced
by the free Hamiltonian is much slower than the Zeno effect induced
by the frequent projective measurements. So in the Zeno limit $\tau\rightarrow0$,
the impact of the free Hamiltonian on the decay of the survival probability
can be neglected. This is also the reason why the free Hamiltonian
does not appear in the effective decay rate of survival probability
\eqref{eq:42}.

\begin{figure*}[!t]
\includegraphics[width=17.2cm]{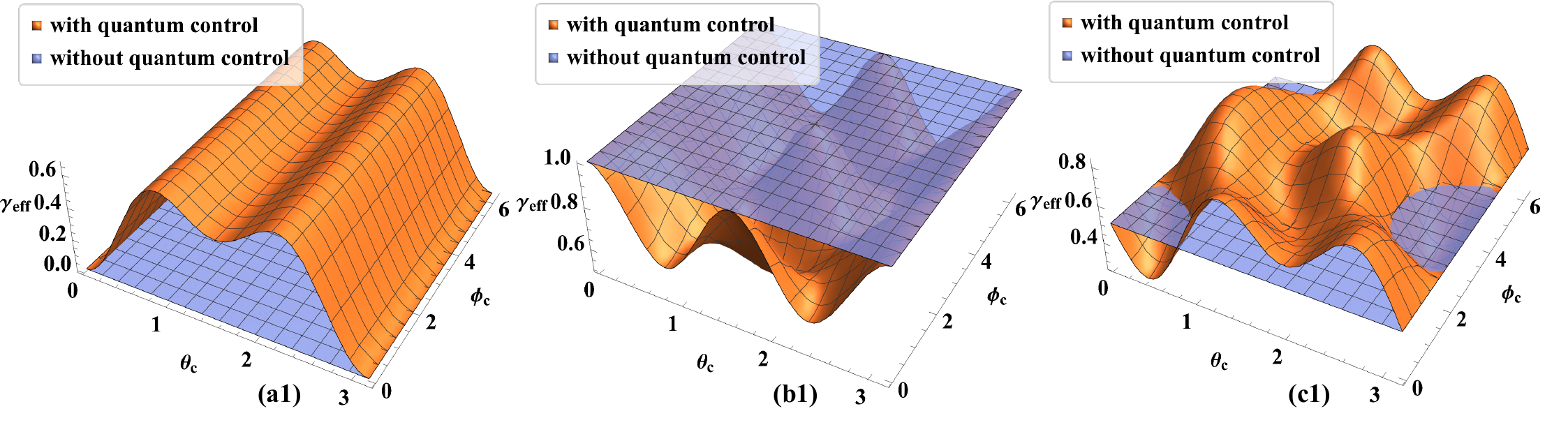}

\includegraphics[width=17.2cm]{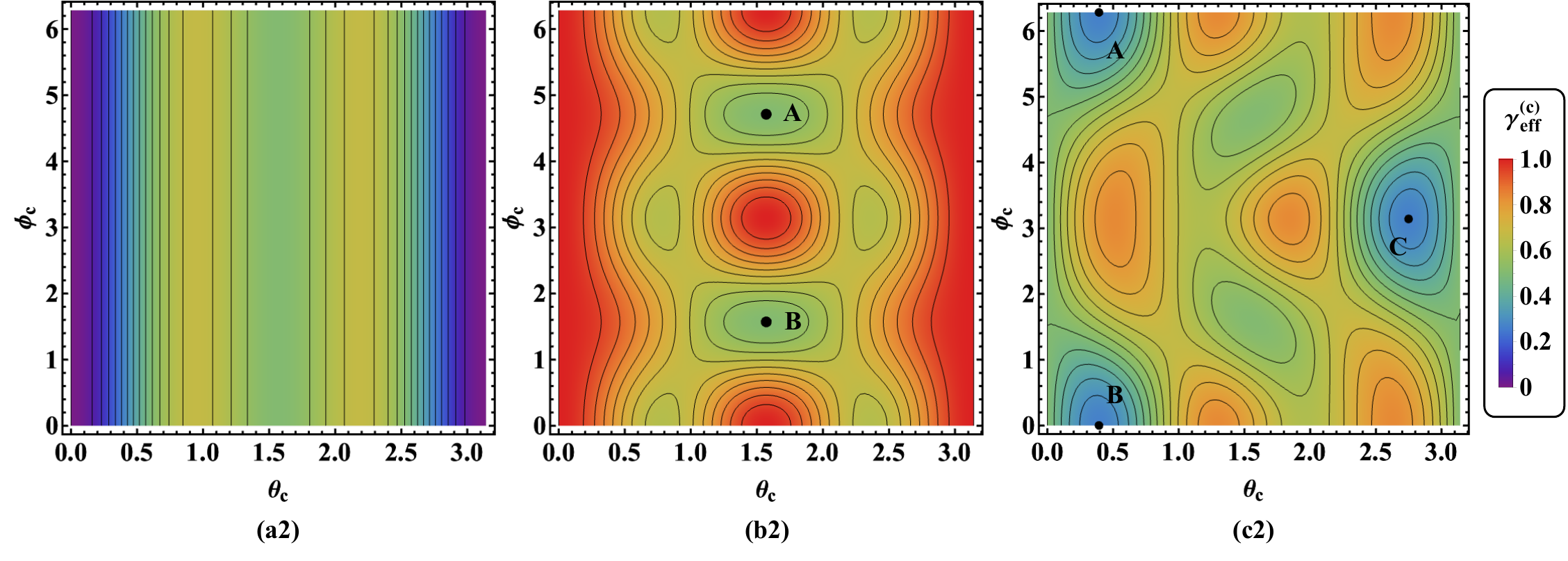}\caption{\label{fig:phaseinitial} Illustration of effective decay rate $\protect\gammaeff$
for different initial states: (a) $\alpha_{\protect\zheng a}=0$ (b)
$\alpha_{\protect\zheng b}=\pi/2$ and (c) $\alpha_{\protect\zheng c}=\pi/4$,
with and without quantum control against dephasing noise. The decay
rate $\protect\gammaeff^{(\protect\noise)}$ without the coherent
control is plotted by the blue surface, and the decay rate $\protect\gammaeff^{(\protect\control)}$
with the coherent control is plotted by the orange surface. When the
initial state is $\alpha=0$, the direction of the optimal quantum
control is parallel to that of the initial state $\protect\thetakong[\protect\control]=0$
or $\pi$, which is a trivial control. When the initial state is $\alpha=\pi/2$,
the directions for the most optimal controls can be obtained as $\left\{ \protect\thetakong[\protect\control]=\pi/2,\protect\ph[\protect\control]=\pi/2,3\pi/2\right\} $,
the effective decay rates under which are indicated by the points
A and B in the subfigure (b2). When the initial state is $\alpha=\pi/4$,
the optimal controls can be found as $\left\{ \protect\thetakong[\protect\control]=\pi/8,\protect\ph[\protect\control]=0;\protect\thetakong[\protect\control]=7\pi/8,\protect\ph[\protect\control]=\pi\right\} $
and and effecive decay rate is indicated by the three annotated points
A, B, C in the subfigure (c2). Parameters: $\beta=0$, $\omega=\pi$
and $\protect\inten=1$.}
\end{figure*}

On the contrary, when the Bloch vector of the initial state stays
on the equator of the Bloch sphere, i.e., $\alpha=\pi/2$, the impact
of the dephasing noise is most significant and the probability for
the system to survive in the initial state is worst. In this case,
any Bloch vector other than those on the equator can suffer less from
the dephasing noise, so any coherent quantum control can improve the
survival probability of the system in the presence of the dephasing
noise though the improvement can differ by different control Hamiltonians.
And similar as above, the impact of the free Hamiltonian is negligible
as we are considering the Zeno limit here.

The initial states along the z axis and on the equator of the Bloch
sphere are the two limiting cases regarding the influence of the dephasing
noise on the survival probability and the extent to which coherent
quantum controls may help. For any other intermediate cases, the initial
states can suffer from the dephasing noise but not as much as those
on the equator of the Bloch sphere, and accordingly coherent quantum
controls can improve the survival probability of system to stay in
the initial states but not as much as for the initial states on the
equator. And actually coherent quantum controls may even worsen the
survival probability if the control Hamiltonian is not chosen properly
in this case.

To gain an intuitive picture of how different choices of the initial
state affect the impact of the dephasing noise and the extent that
coherent quantum controls can improve the survival probability, the
effective decay rates of the survival probability in the above three
cases are depicted in Fig.$\ $\ref{fig:phaseinitial} for three typical
initial states,
\begin{equation}
\begin{aligned}|\psi_{\zheng a}\rangle & =|0\rangle,\\
|\psi_{\zheng b}\rangle & =\frac{1}{\sqrt{2}}\left(|0\rangle+|1\rangle\right),\\
|\psi_{\zheng c}\rangle & =\cos\frac{\text{\ensuremath{\pi}}}{8}|0\rangle+\sin\frac{\text{\ensuremath{\pi}}}{8}|1\rangle,
\end{aligned}
\end{equation}
which belong to the three different cases respectively.

Fig. \ref{fig:phaseinitial} (a1) and (a2) depict the effective decay
rate for the initial state $|\psi_{\zheng a}\rangle$ with $\alpha=0$,
lying along the $z$-axis of the Bloch sphere. And it can be seen
that $\gammaeff^{(\control)}\geq\gammaeff^{(\noise)}$, i.e., any
coherent control can only induce the decay of survival probability
or keep it unchanged at most, since the initial state is already in
the most favorable direction which is free from the impact of dephasing
and any coherent control scheme may not reduce the decay in this case.
As a contrast, Fig. \ref{fig:phaseinitial} (b1) and (b2) depict the
decay rate for the initial state $|\psi_{\zheng b}\rangle$ with $\alpha=\pi/2$,
$\beta=0$, lying on the equator of the Bloch sphere. It can be seen
from the figure that $\gammaeff^{(\control)}\leq\gammaeff^{(\noise)}$,
i.e., any control Hamiltonian can improve the survival probability
or keep it unchanged at least, since the initial state experiences
the most severe impact of the dephasing noise and thus any coherent
control scheme cannot do worse than without the control. Fig. \ref{fig:phaseinitial}
(c1) and (c2) depict the intermediate case for the initial state $|\psi_{\zheng c}\rangle$
with $\alpha=\pi/4$, $\beta=0$, lying along a direction between
the $z$ axis and the equator of the Bloch sphere. It can be observed
that some choices of the control Hamiltonian can improve the decay
of survival probability while the others may worsen it, as there exist
both directions that suffer more or less from the dephasing noise
on the Bloch sphere and the control Hamiltonian may rotate the system
to either of them in this case. As shown by Fig. \ref{fig:phaseinitial},
different directions of the control Hamiltonian have different capabilities
to improve the survival probability given the initial state of the
system, so in the following our objective is to find the optimal direction
of the control Hamiltonian that minimize the effective decay rate
$\gammaeff^{(\control)}$ to protect the Zeno effect.

According to the general variation equation \eqref{eq:81} for a two-level
system along with the positive definiteness of the Hessian matrix
\citep{bazaraa2013nonlinear} for $\gammaeff^{(\optimal)}$ with respect
to direction parameters $\thetakong[\control]$ and $\ph[\control]$
of the control Hamiltonian to ensure the minimization (not the maximization)
of the decay rate $\gammaeff^{(\control)}$, we have the following
optimization equations for $\thetakong[\control]$ and $\ph[\control]$,
\begin{equation}
\begin{aligned}\partial_{\thetakong[\control]}\gammaeff^{(\control)} & =0,\\
\partial_{\ph[\control]}\gammaeff^{(\control)} & =0,\\
A>0,\;C & >0,\\
AC-B^{2} & <0,
\end{aligned}
\label{eq:37}
\end{equation}
where $A=\partial_{\thetakong[\control]}^{2}\gammaeff^{(\control)}$,
$C=\partial_{\ph[\control]}^{2}\gammaeff^{(\control)}$ and $B\equiv\partial_{\thetakong[\control]}\partial_{\ph[\control]}\gammaeff^{(\control)}$.

One can obtain the optimal directions of the control Hamiltonian in
the presence of dephasing noise given the Bloch vector $\boldsymbol{r_{0}}$
of the initial state of the two-level quantum system by solving Eq.
\eqref{eq:37},
\begin{equation}
\begin{cases}
\thetakong[\control]=\frac{\alpha}{2},\ph[\control]=\beta, & \arccos\left(\frac{1}{3}\right)<\alpha\leq\pi,\\
\thetakong[\control]=\frac{\pi+\alpha}{2},\ph[\control]=\beta, & 0\leq\alpha<\arccos\left(-\frac{1}{3}\right),\\
\thetakong[\control]=\frac{\pi}{2},\ph[\control]=\beta+\frac{\pi}{2}.
\end{cases}\label{eq:23}
\end{equation}

Note that the solutions in Eq. \eqref{eq:23} include all the local
optimal points for the direction of the control Hamiltonian. To find
the global optimal point for the control Hamiltonian, one needs to
substitute this solution into the effective decay rate $\gammaeff^{(\control)}$
\eqref{eq:42} and compare the results corresponding to the three
scenarios in Eq. \eqref{eq:23}. The global minimum effective decay
rate turns out to be
\begin{equation}
\gammaeff^{(\optimal)}=\begin{cases}
-\frac{\inten}{16}\left(-7+4\cos\alpha+3\cos2\alpha\right), & 0\leq\alpha<\alpha_{0},\\
\frac{\inten}{2}, & \alpha_{0}\leq\alpha<\alpha_{1},\\
-\frac{\inten}{4}\cos^{2}\alpha\left(-5+3\cos\alpha\right), & \alpha_{1}\leq\alpha\leq\pi,
\end{cases}\label{eq:33}
\end{equation}
where $\alpha_{0}\equiv2\arccos\sqrt{2}$ , $\alpha_{1}=\pi-\alpha_{0}$
and $\inten$ is the noise strengh introduced in the master equation
\eqref{eq:meq}.

It can be observed that the optimal effective decay rate $\gammaeff^{(\optimal)}$
with the coherent quantum control and the decay rate $\gammaeff^{(\noise)}$
without any control are both independent of the azimuthal angle $\beta$
of the Bloch vector of the initial state. This characteristic arises
from the rotational symmetry of dephasing noise about the $z$-axis.
\begin{figure}
\includegraphics[width=8.6cm]{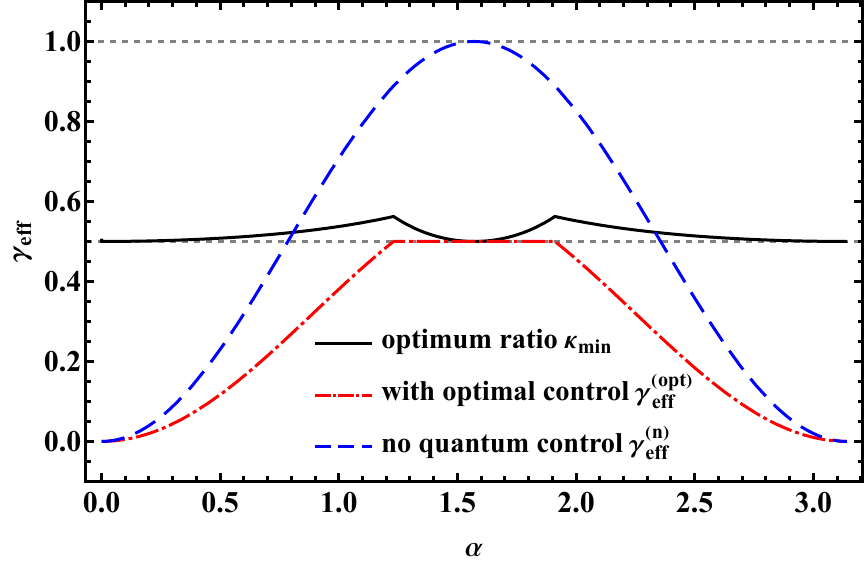}\caption{\label{fig:3}Illustration of effective decay rates of the survival
probability with or without the coherent control in the presence of
dephasing noise for different initial states. The decay rate $\protect\gammaeff^{(\protect\noise)}$
without the coherent control is plotted by the blue dashed line, and
the minimum decay rate $\protect\gammaeff^{(\protect\optimal)}$ with
the optimized coherent control is plotted by the red dot-dashed line.
The ratio $\kappa_{\mathrm{min}}=\protect\gammaeff^{(\protect\optimal)}/\protect\gammaeff^{(\protect\noise)}$
is also plotted by the black solid line, which shows the stability
of the optimization performance of this coherent control scheme over
different initial states of the system. Parameters: $\omega=\pi$
and $\protect\inten=1$.}
\end{figure}

\begin{figure}
\includegraphics[width=8.6cm]{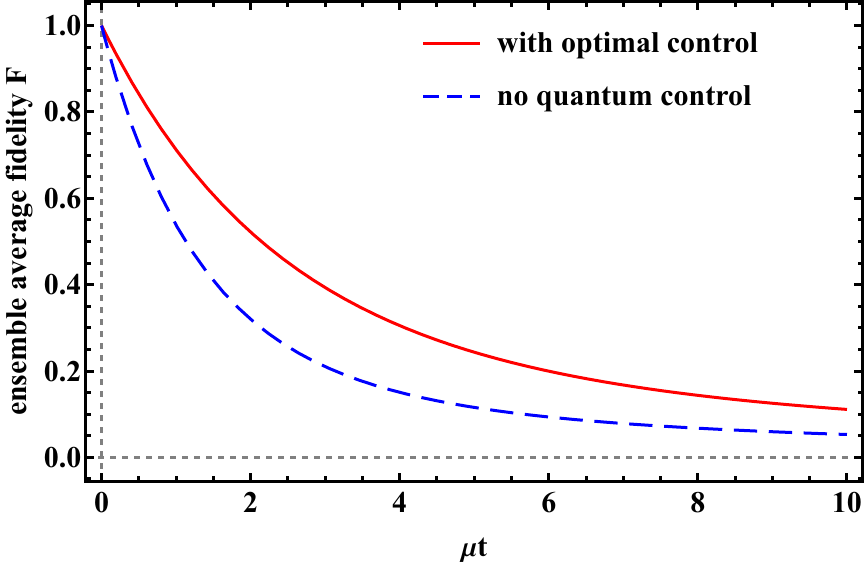}

\caption{\label{fig:phaseave}Plots of the ensemble average fidelity $F$ with
respect to $\protect\inten t$ in the presence of dephasing noise
with the optimal coherent control scheme and without any quantum control,
respectively.}
\end{figure}

\begin{figure*}
\includegraphics[width=17.2cm]{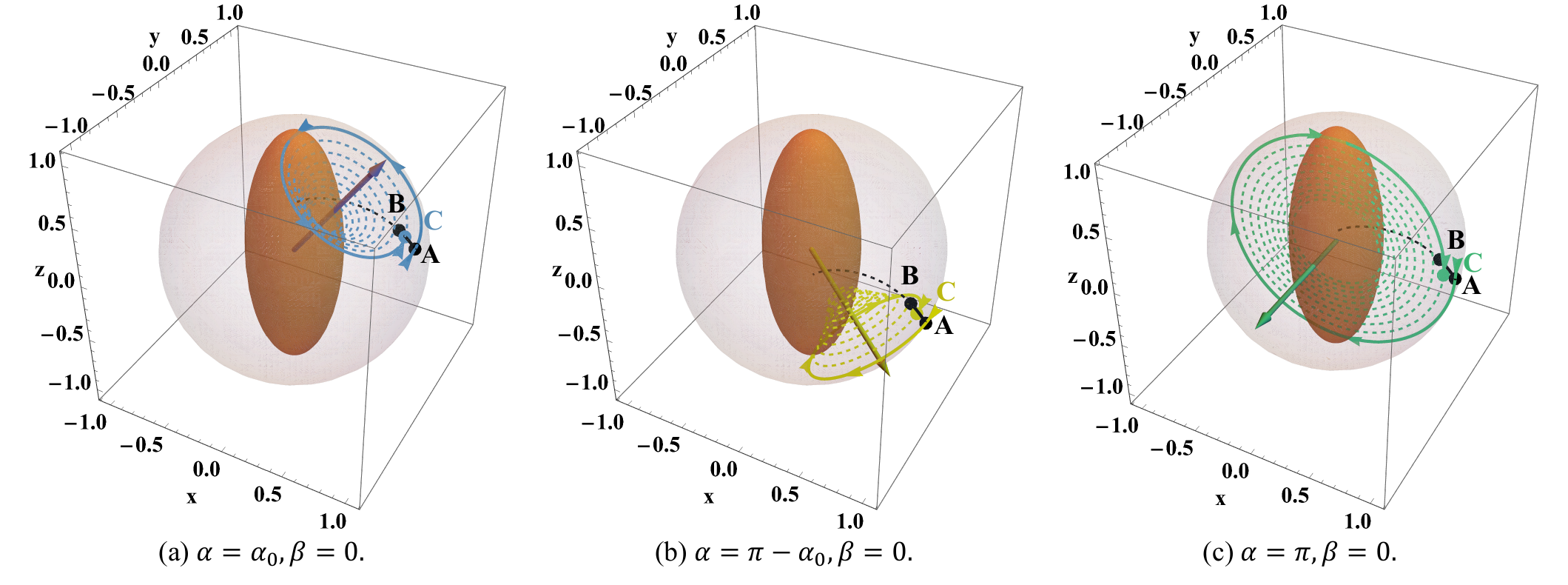}\caption{\label{fig:2}Illustration for the evolution paths of a two-level
quantum system initially prepared on different pure states with and
without the protection by the optimal coherent control between two
consecutive projective measurements in the presence of dephasing noise.
In each subfigure, the gray, the translucent sphere is the Bloch sphere
consisting of all density matrices of a single qubit, the orange,
the translucent ellipsoid is the set of all density matrices after
the disturbance of the dephasing noise on the system, and the arrow
indicates the direction of optimal control Hamiltonian. The point
A represents the initial state, and the points B and C represent the
final states at the end of each time interval without and with the
Hamiltonian control respectively. The black lines depict the evolution
paths of the system without quantum controls, whereas the colored
lines represent the evolution paths engineered by the optimal controls.
The solid arcs of the colored lines depict the actual evolution paths
between two consecutive projective measurements, while the dashed
arcs of the colored lines depict the following evolution paths if
the evolution is not interrupted by the projective measurement. Typical
initial states of the quantum system are chosen for the three different
regimes given by Eq. \eqref{eq:33} respectively, (a) $\alpha\protect\leq\alpha_{0}$
(b) $\alpha\protect\geq1-\alpha_{0}$ and (c) $\alpha_{0}<\alpha<1-\alpha_{0}$.
Parameters: $\protect\hami[0]=\sigma_{z},\,\beta=0,\,\mu=1,\,\omega=\pi,\,\tau=0.01.$}
\end{figure*}
The relations between the decay rates $\gammaeff^{(\optimal)}$, $\gammaeff^{(\noise)}$
and the polar angle $\alpha$ of the Bloch vector of the initial state
as well as the optimal ratio $\kappa$ for different initial states
are depicted in Fig.$\ $\ref{fig:3}. In the figure, one can observe
that the optimized effective decay rate $\gammaeff^{(\optimal)}$
always remains lower than the decay rate without control $\gammaeff^{(\noise)}$,
indicating the effectiveness of the coherent control scheme. It can
also be observed from the figure that the coherent control scheme
is robust against the change of the polar angle $\alpha$ as the optimized
ratio $\kappa$ has only minor fluctuation over the whole range of
$\alpha$, achieving the minimum value $1/2$ near the poles ($\alpha=0$
or $\pi$) of the Bloch sphere where the influence of the dephasing
noise is negligible and on the equator ($\alpha=\pi/2$) where the
influence of the dephasing noise is most significant. The optimization
performance is poorest at $\alpha=\alpha_{0}$ and $\pi-\alpha_{0}$,
where $\kappa=9/16$. So the ratio $\kappa$ changes only slightly
over the range of $\alpha$.

To characterize the overall performance of the above optimized coherent
quantum control scheme over different initial states of the system,
we consider the ensemble average fidelity $F$ \eqref{eq:ensemblefidlity}
over uniformly distributed initial states, which turns out to be
\begin{equation}
F\left(t\right)=\frac{1}{4\pi}\int_{0}^{2\pi}d\beta\int_{0}^{\pi}d\alpha e^{-\gammaeff t}\sin\alpha,\label{eq:34}
\end{equation}
for a two-level system. The spherical integral is due to the distribution
of initial states on the Bloch sphere with a radius $|\boldsymbol{r_{0}}|=1$,
and $\frac{1}{4\pi}$ is the normalization coefficient.

By substituting Eqs.$\ $\eqref{eq:32} and \eqref{eq:33} into \eqref{eq:34},
one can obtain the decay of the ensemble average fidelity $F$ with
respect to $\inten t$ in both the control-free and optimally controlled
cases, as shown in Fig.$\ $\ref{fig:phaseave}. It is evident from
the figure that the ensemble average fidelity $F$ with the optimal
coherent control is always greater than that in the control-free scenario,
indicating a slower decay of the survival probability with the optimal
coherent control given the same noise intensity $\inten$. This observation
explicitly demonstrates the effectiveness of the above coherent quantum
control strategy in protecting the quantum Zeno effect against the
dephasing noise.

It is helpful to pause and ponder the physical mechanism behind the
optimization effect of the above coherent control scheme. For the
dephasing noise, the $z$ axis is the direction that is not disturbed
by the noise, so it would be beneficial to rotate a quantum state
towards the $z$ axis during the evolution by quantum control to reduce
the influence of the dephasing noise. As the Hamiltonian control is
a coherent control scheme which preserves the purity of a quantum
system, one actually wants to rotate the quantum system towards the
$|0\rangle$ or $|1\rangle$ state, i.e., the north or south pole
of the Bloch sphere.

This is indeed what the above Hamiltonian control scheme does, as
illustrated by Fig. \ref{fig:2}. From the three colored lines in
Fig. \ref{fig:2} which represent the evolution paths of the quantum
system with the optimal control Hamiltonians, it can be seen that
the Hamiltonian control drags the quantum system towards the $|0\rangle$
state (as the initial state is chosen to be in the upper semisphere
in the figure which is closer to $|0\rangle$) and then turns it back
to the vicinity of the initial state (as the purpose is to preserve
the initial state), which is the physical significance of the condition
\eqref{eq:69}. The joint effect of the Hamiltonian control and the
dephasing noise is to approximately rotate the state of quantum system
between the initial state and the north/south pole of Bloch sphere
along a spiral path towards the $z$ axis, realizing the decrease
of decay caused by the dephasing noise. As a contrast, the quantum
system evolves along the black paths without quantum control which
decays to the $z$ axis faster.

It is worth noting that the system cannot perfectly return to the
initial state at the end of each cycle in spite of the Hamiltonian
control, due to the existence of non-zero first-order term in the
survival probability which can be lowered by the coherent control
scheme but not eliminated. But it can be seen from Fig. \ref{fig:2}
that the distance between the initial state (point A) and the final
state with the quantum control (point C) is always shorter than that
between the initial state and the final state without the quantum
control (point B), indicating the effectiveness of the above quantum
control scheme.

It is also worth mentioning that while the unitary noise considered
in this subsection is the dephasing noise, the above optimal control
scheme can be generalized for arbitrary unitary noise, since the effect
of a unitary noise rather than the dephasing is equivalent to a new
free Hamiltonian with the dephasing noise in a rotated picture which
can be included above due to the arbitrariness of free Hamiltonian
assumed in the above study.

\subsubsection{Amplitude damping}

The amplitude damping is another typical noise on a two-level system,
usually describing the energy loss from a quantum system, known as
energy dissipation. The noise coefficient matrix \eqref{eq:gammatrix}
of amplitude damping is
\begin{equation}
\Gamma_{\zheng{ad}}=\frac{\inten}{4}\left(\begin{array}{ccc}
1 & -i & 0\\
i & 1 & 0\\
0 & 0 & 0
\end{array}\right),\label{eq:gad}
\end{equation}
and the master equation for the evolution of the system with a free
Hamiltonian and the amlitude damping noise is
\begin{equation}
\frac{d\rot}{dt}=-i\left[\hami[0],\rot\right]+\inten\huaD[\sigma_{-}]\rot,\label{eq:36}
\end{equation}
where $\huaD[\sigma_{-}]\rot=\sigma_{-}\rot\sigma_{+}-\frac{1}{2}\left\{ \sigma_{+}\sigma_{-},\rot\right\} $
and $\rot$ is the density matrix of the system at time $t$. When
a control Hamiltonian $g\hami[\control]$ is applied on the system,
the master equation becomes
\begin{equation}
\frac{d\rot}{dt}=-i\left[\hami[0]+g\hami[\control],\rot\right]+\inten\huaD[\sigma_{-}]\rot.
\end{equation}

\begin{figure}
\includegraphics[width=4.3cm]{1a}\includegraphics[width=4.3cm]{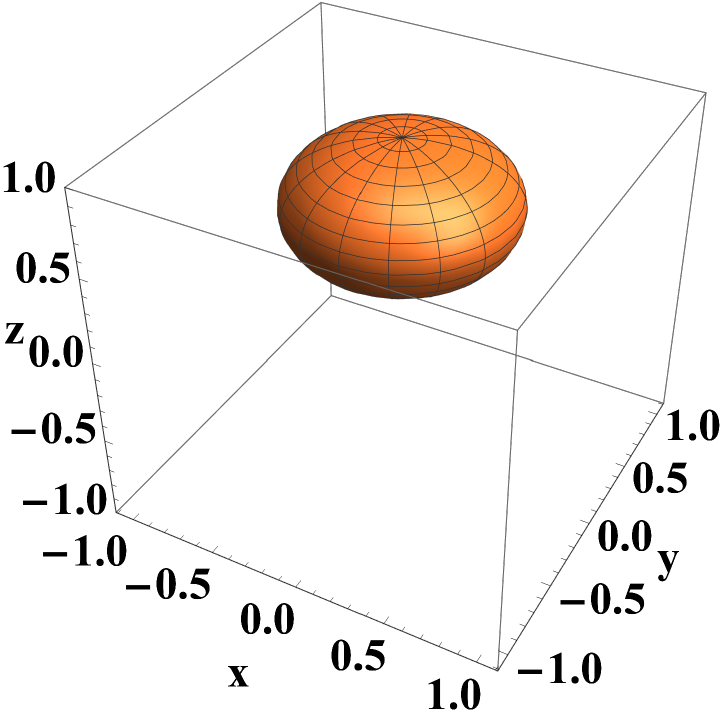}

\caption{\label{fig:7}The effect of the amplitude damping noise on a two-level
system. The amplitude damping noise compresses the Bloch sphere towards
the north pole, i.e., the quantum state $|0\rangle$, which is the
stationary state of the amplitude damping noise. Parameter: $\protect\inten t=1$.}
\end{figure}

The effect of the amplitude damping noise on a two-level system is
depicted in Fig. \ref{fig:7}. It shows that the amplitude damping
noise compresses the Bloch sphere towards the north pole, i.e., the
state $|0\rangle$ which is the unique stationary state of the amplitude
damping noise. The compression is rotationally symmetric around the
$z$ axis as the amplitude damping noise affects the $\sigma_{x}$
and $\sigma_{y}$ components of all density matrices uniformly, but
in contrast to the dephasing noise discussed above, it is not symmetric
about the equatorial plane as the amplitude damping noise also changes
the $\sigma_{z}$ component of a density matrix and this change varies
with the $\sigma_{z}$ component of the density matrix.

When the projective measurement is performed sufficiently frequently
and the condition \eqref{eq:69} to preserve the initial state is
satisfied in the Zeno limit $\tau\rightarrow0$, by substituting $\boldsymbol{r_{0}}$
\eqref{eq:69} and $\Gamma_{\mathrm{ad}}$ \eqref{eq:gad} into the
Eqs.$\ $\eqref{eq:18} and \eqref{eq:66}, one can obtain the effective
dacay rate of survival probability without the Hamiltonian control
as
\begin{equation}
\gammaeff^{(\noise)}=\frac{\inten}{8}\left(3-4\cos\alpha+\cos2\alpha\right)=\inten\sin^{4}\frac{\alpha}{2},\label{eq:35}
\end{equation}
and the effective decay rate with the control Hamiltonian as
\begin{equation}
\begin{aligned}\gammaeff^{(\control)}= & \frac{\inten}{512}\Big\{178+12\cos2\left(\alpha-\thetakong[\control]\right)+\cos2(\alpha-\Delta)\\
 & +\cos2(\alpha+\Delta)-256\cos\alpha\cos^{2}\thetakong[\control]+8\cos2\thetakong[\control]\\
 & +6\cos4\thetakong[\control]+2\cos2\alpha(11+6\cos2\thetakong[\control]+9\cos4\thetakong[\control])\\
 & +2\cos2\Delta(-1+2(4\cos2\thetakong[\control]-3\cos4\thetakong[\control])\sin^{2}\alpha)\\
 & +4\Big[-32\cos\Delta\sin\alpha+\big(4\cos\Delta(1+3\cos2\thetakong[\control])\\
 & \ -3\big)\sin2\alpha\Big]\sin2\thetakong[\control]\Big\},
\end{aligned}
\label{eq:38}
\end{equation}
where $\Delta=\beta-\ph[\control]$.

\begin{figure*}
\includegraphics[width=17.2cm]{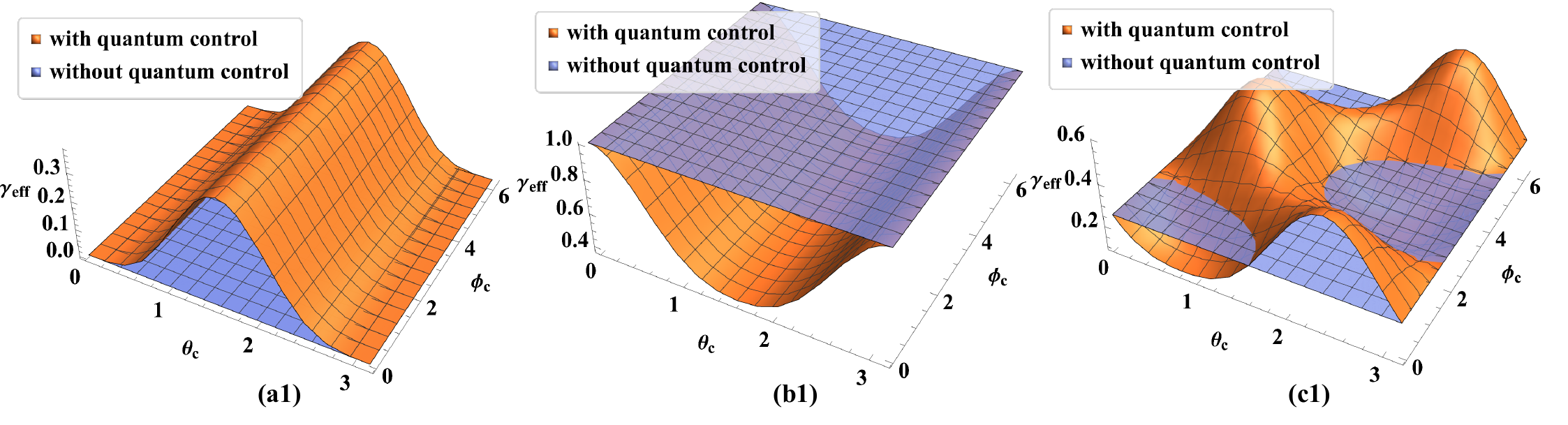}

\includegraphics[width=17.2cm]{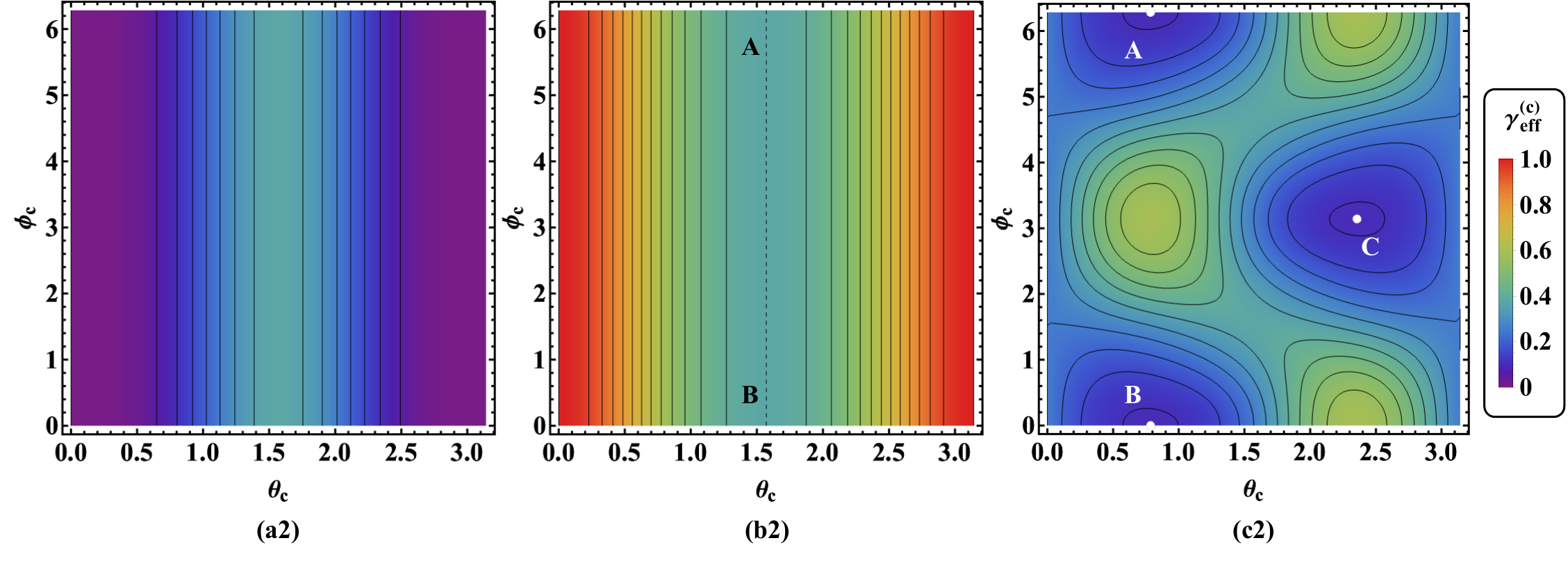}\caption{\label{fig:9}Illustration of effective decay rates $\protect\gammaeff$
for different initial states (a) $\alpha_{\protect\zheng a}=0$ (b)
$\alpha_{\protect\zheng b}=\pi$ and (c) $\alpha_{\protect\zheng c}=\pi/2$
in the presence of amplitude damping noise, with the Hamiltonian controls
of all possible directions versus without a Hamiltonian control. The
decay rates $\protect\gammaeff^{(\protect\noise)}$ without the controls
are plotted by the blue surface, and the decay rates $\protect\gammaeff^{(\protect\control)}$
with the coherent control are plotted by the orange surface. When
the initial state is $\alpha=0$, the optimal quantum control direction
is parallel to the initial state direction $\protect\thetakong[\protect\control]=0$
or $\pi$, which is trivial control. When the initial state is $\alpha=\pi$,
the direction of the most optimal control can be identified as $\left\{ \protect\thetakong[\protect\control]=\pi/2\right\} $,
and the corresponding effective decay rates are represented by the
dashed line AB in the subfigure (b2). When the initial state is $\alpha=\pi/2$,
the optimal controls can be found as $\{\protect\thetakong[\protect\control]=\pi/4,\protect\ph[\protect\control]=0;\protect\thetakong[\protect\control]=3\pi/4,\protect\ph[\protect\control]=\pi\}$
and the effective decay rates are represented by the three annotated
points A, B and C in the subfigure (c2). Parameters: $\beta=0$, $\omega=\pi$
and $\protect\inten=1$.}
\end{figure*}

From the result of $\gammaeff^{(\noise)}$ \eqref{eq:35} without
Hamiltonian control, it's apparent that the decay rate of the survival
probability induced by the amplitude damping noise is independent
of the azimuthal angle $\beta$ and solely depends on the polar angle
$\alpha$ between the initial state and the $z$ axis. This is in
accordance with the rotational symmetry of the compression effect
of the amplitude damping noise around the $z$ axis shown in Fig.
\ref{fig:7}. However, an additional term, $\cos\alpha$, is introduced
in $\gammaeff^{(\noise)}$ \eqref{eq:35}, which is not symmetric
about $\alpha=\pi/2$, so the decay rate is not symmetric about the
equatorial plane and the decay increases with $\alpha$, which also
agrees with Fig.$\ $\ref{fig:7}. The north pole of the Bloch sphere,
i.e. the state $|0\rangle$, is a stationary state of the amplitude
damping noise, so it remains unaffected by the amplitude damping noise,
and no decay of the survival probability occurs when the system is
initially in this state. On the contrary, the south pole of the Bloch
sphere, i.e., the state $|1\rangle$, is the state whose Bloch vector
is compressed most by the amplitude damping noise, so when the initial
state of the system resides at the south pole, the survival probability
decays most significantly with time.

When coherent quantum control is applied on the system, similar as
the dephasing noise discussed above, the impact of the amplitude damping
noise differs with the initial state of the two-level system. When
the initial state of the system is $|0\rangle$, i.e., $\alpha=0$,
the north pole of the Bloch sphere, the amplitude damping noise does
not change the system as $|0\rangle$ is the stationary state of the
amplitude damping noise and the survival probability does not decay.
So applying any control Hamiltonian on the system can only induce
decay on the survival probability. On the contrary, when the system
is initially in the state $|1\rangle$, i.e., $\alpha=\pi$, the south
pole of the Bloch sphere, the system suffers the most disturbance
from the amplitude damping noise and the survival probability decays
fastest. So in this case, introducing any control Hamiltonian to the
system can help slow the decay of survival probability. When the system
initially stays at any state other than $|0\rangle$ or $|1\rangle$,
a control Hamiltonian may decrease or increase the decay rate of the
survival probability, as one can always find another state that is
better or worse than the initial state in suffering the amplitude
damping noise, which is an intermediate case between the states $|0\rangle$
and $|1\rangle$.

To visualize the effects of the Hamiltonian controls on different
initial states of the system in the presence of the amplitude damping
noise, the effective decay rate of the survival probability $\gammaeff^{(\control)}$
\eqref{eq:38} with all possible directions of the control Hamiltonian
is plotted in Fig. \ref{fig:9} for three typical initial states of
the system,
\begin{equation}
\begin{aligned}|\psi_{\zheng a}\rangle & =|0\rangle,\\
|\psi_{\zheng b}\rangle & =|1\rangle,\\
|\psi_{\zheng c}\rangle & =\frac{1}{\sqrt{2}}\left(|0\rangle+|1\rangle\right),
\end{aligned}
\end{equation}
which falls into the three different categories of the states discussed
above respectively.

Figs. \ref{fig:9} (a1) and (b1) depict the effective decay rate for
the state $|\psi_{\zheng a}\rangle$, i.e., $\alpha=0$. It shows
that $\gammaeff^{(\control)}\geq\gammaeff^{(\noise)}$ for all directions
of the control Hamiltonian, i.e., any coherent control can only induce
decay in the survival probability or keep it unchanged at most, as
the system is not affected by the amplitude damping noise and the
survival probability cannot benefit from the control Hamiltonian in
any direction in this case. Figs. (a2) and (b2) depict the case for
the state $|\psi_{\zheng b}\rangle$, i.e., $\alpha=\pi$, and show
that $\gammaeff^{(\control)}\leq\gammaeff^{(\noise)}$ for all directions
of the control Hamiltonian, i.e., any coherent control can help slow
the decay of survival probability or keep it unchanged at least, as
$|1\rangle$ is the state most adversely affected by the amplitude
damping noise, and thus the Hamiltonian control in an arbitrary direction
can mitigate this situation. Figs. (a3) and (b3) consider the intermediate
case with the initial state $|\psi_{\zheng c}\rangle$, i.e., $\alpha=\pi/2$,
$\beta=0$, demonstrating that the possibilities for coherent quantum
controls to reduce or increase the decay rate of the survival probability
exist simultaneously, as both states that are less or more disturbed
by the amplitude damping noise exist on the Bloch sphere in this case.

As the improvement in the decay rate of the survival probability differs
among different directions of the control Hamiltonian, it is desirable
to find the lowest decay rate by optimizing the control Hamiltonian
over all possible directions.

\begin{figure}
\includegraphics[width=8.6cm]{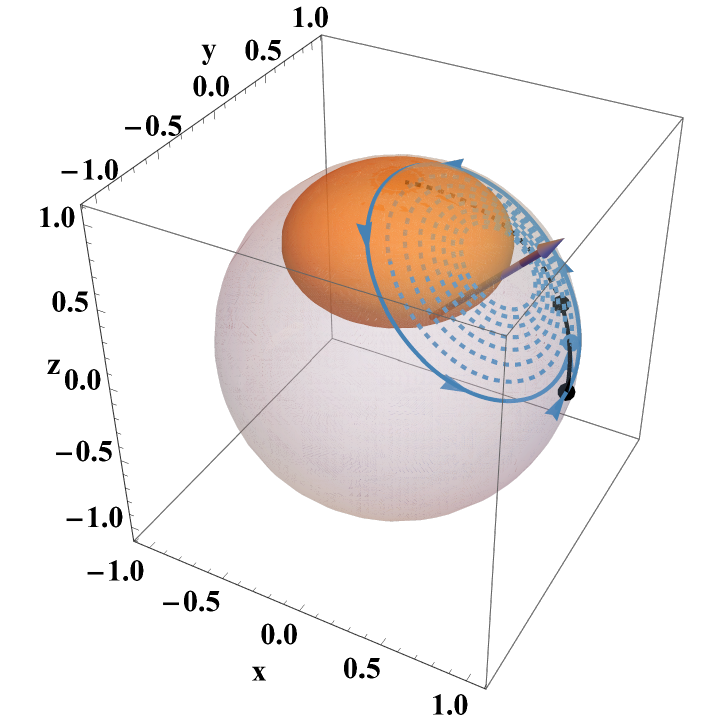}

\caption{\label{fig:12}The evolution paths of a two-level quantum system initially
prepared on a pure state with and without the protection of the optimal
coherent control between two consecutive projective measurements.
The gray, translucent unit sphere represents the set of all density
matrices of a two-level system. The deformed, translucent, orange
ellipsoid is set of all final density matrices transformed by the
amplitude damping noise on the system, and the arrow indicates the
direction of optimal coherent control. The point A represents the
initial state of the system, and the points B and C represent the
respective final states after the evolution between two consecutive
measurements without and with the optimal coherent quantum control.
The black line depicts the evolution path without the control, and
the blue line represents the evolution path engineered by the optimal
control. The solid arc of the blue line denotes the actual evolution
path between two consecutive projective measurements, while the dashed
arc of the blue line denotes the future evolution path if the evolution
is not interrupted by the repetitive projective measurements. Parameters:
$\protect\hami[0]=\sigma_{z},\mu=1,\omega=\pi,\tau=0.25.$}
\end{figure}

Substituting Eq.$\ $\eqref{eq:38} into Eq.$\ $\eqref{eq:37}, one
can work out the optimal control for the amplitude damping noise,
which reveals that for an initial state with a Bloch vector $\boldsymbol{r_{0}}$
\eqref{eq:r0} the effective decay rate reaches the minimum when
\begin{equation}
\thetakong[\control]=\frac{\alpha}{2},\;\ph[\control]=\beta.\label{eq:39}
\end{equation}
The evolution trajectory of the Bloch vector of the system with the
control Hamiltonian in the optimal direction \eqref{eq:39} is plotted
in Fig. \ref{fig:12}.

Similar to the case of dephasing noise, the mechanism of the Hamiltonian
control against the amplitude damping noise can be understood from
the evolution paths of the quantum system with and without the optimal
control, which is illustrated in Fig. \ref{fig:12}. The evolution
of the system without control is plotted by the black path of Fig.
\ref{fig:12} which shows that the quantum system would directly approach
the ground state $\rig[0]$ under the influence of amplitude damping
noise in this case. However, the rotation under the combined influence
of coherent control and amplitude damping noise, shown by the blue
path of Fig. \ref{fig:12}, suggests that the effect of the coherent
control drags the state towards $|0\rangle$, i.e., the north pole
of Bloch sphere, which is the state least influenced by the amplitude
damping noise, and then turns it back to the vicinity of the initial
state. The result of such a Hamiltonian-controlled evolution ensures
the distance between the initial state (point A) and the final state
with the control (point C) shorter than that between the initial state
and the final state without the control (point B) and thus achieves
the purpose of delaying the decay of the quantum system.

\begin{figure}
\includegraphics[width=8.6cm]{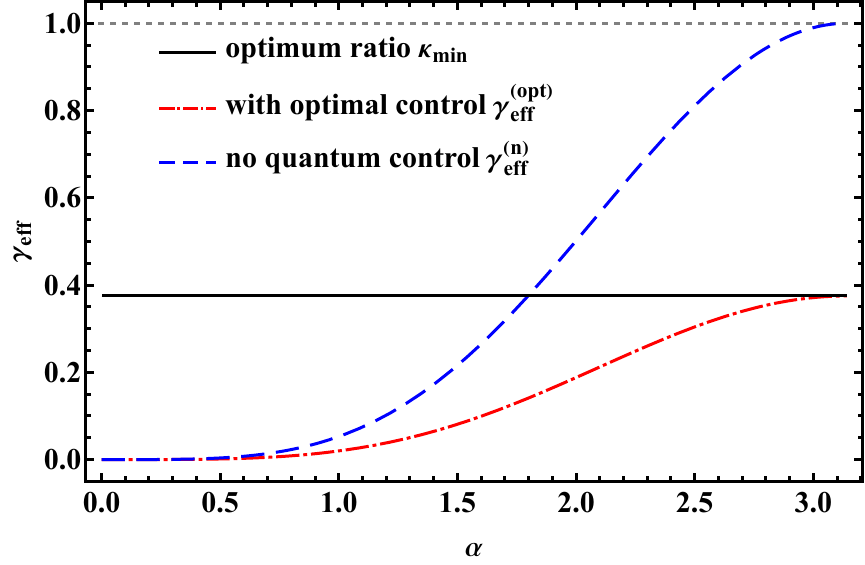}\caption{\label{fig:8}Illustration of effective decay rates of the survival
probability with or without the coherent control in the presence of
amplitude damping noise for different initial states. The decay rate
$\protect\gammaeff^{(\protect\noise)}$ without the coherent control
is plotted by the blue dashed line, and the minimum decay rate $\protect\gammaeff^{(\protect\optimal)}$
with the optimized coherent control is plotted by the red dot-dashed
line. The ratio $\kappa=\protect\gammaeff^{(\protect\optimal)}/\protect\gammaeff^{(\protect\noise)}$
is also plotted by the black solid line, showing the stability of
the optimization performance of this coherent control scheme over
different initial states of the system. Parameters: $\omega=\pi$
and $\protect\inten=1$.}
\end{figure}

With the optimal coherent control scheme in Eq.$\ $\eqref{eq:39},
the effective decay rate of survival probability induced by the amplitude
damping noise for any arbitrary initial state can reach its minimum,
which turns out to be
\begin{equation}
\gammaeff^{(\optimal)}=\frac{3}{8}\inten\sin^{4}\frac{\alpha}{2},\label{eq:40}
\end{equation}
implying that the ratio $\kappa$ \eqref{eq:67} optimized by the
coherent control scheme is the same for all initial pure states, which
is $\kappa=\gammaeff^{(\optimal)}/\gammaeff^{(\noise)}=3/8$. This
demonstrates the effectiveness and stability of this optimized coherent
quantum control approach. The optimal effective decay rate with the
optimal coherent control $\gammaeff^{(\optimal)}$ and without any
control $\gammaeff^{(\noise)}$ are both independent of the azimuthal
angle $\beta$ of the initial state due to the rotational symmetry
of amplitude damping noise. Their relations with the polar angle $\alpha$
of the initial state as well as the best ratio $\kappa$ is depicted
for different initial states in Fig.$\ $\ref{fig:8}.

\begin{figure}
\includegraphics[width=8.6cm]{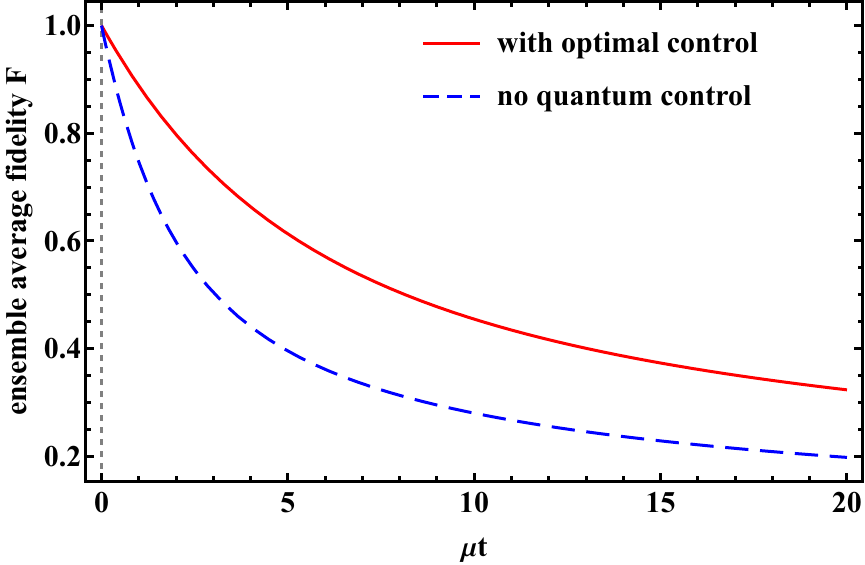}

\caption{\label{fig:amplitudeaverage}The ensemble average fidelity $F$ with
respect to $\protect\inten t$ in the presence of amplitude damping
noise with or without the optimal coherent quantum control scheme
respectively.}
\end{figure}

If the above optimization approach of the coherent quantum control
is applied to each initial state, one can obtain the decay of the
ensemble average fidelity $F$ \eqref{eq:ensemblefidlity} with respect
to $\inten t$ without any quantum control and with the optimal coherent
control by substituting Eqs.$\ $\eqref{eq:35} and \eqref{eq:40}
into Eq.$\ $\eqref{eq:34}. The ensemble average fidelity is plotted
in Fig.$\ $\ref{fig:amplitudeaverage}. It is evident from this figure
that the ensemble average fidelity $F$ with the optimal coherent
control is always greater than than that without any quantum control,
indicating, a decrease in the decay rate of the survival probability
by the optimal coherent control. This observation manifests the validity
of employing this coherent control scheme to protect the survival
probability against the amplitude noise.

Finally, we would like to remark that with the development of quantum
technologies in recent years, two-level quantum systems as well as
the quantum operations on them have been realized in a variety of
physical systems with high precisions, e.g., quantum dots \citep{burkard2023semiconductor},
ion traps \citep{bruzewicz2019trappedion}, superconducting quantum
circuits \citep{blais2021circuit}, etc. We refer the readers to Ref.
\citep{chen2006quantum}for a comprehensive review of physical systems
that can realize two-level systems and the relevant quantum operations.
The coherent controls proposed in the current scheme are essentially
unitary rotations of two-level systems on the Bloch sphere, so they
can also be implemented on those physical systems.

\section{Conclusion\label{sec:CONCLUSION}}

In this work, we consider the quantum Zeno effect in the presence
of noise and study the survival probability that a general quantum
system stays in its initial state by repetitive projective measurements
in this situation. Starting from the master equation with general
dissipative terms, we discuss the physical mechanism underlying the
vanishing of the quantum Zeno effect and the decay of the survival
probability. In order to suppress the influence of the noise, a coherent
control scheme with a strong Hamiltonian is introduced to the quantum
system. As the noise induces a nonzero first-order term in the expansion
of the survival probability of the initial state which leads the decay
of the survival probability, a detailed analysis shows that coherent
quantum control with a Hamiltonian as strong as the frequency of the
projective measurements can reduce the decay rate of the survival
probability. The effective decay rate of the survival probability
with the coherent quantum control is obtained, and the conditions
on the control Hamiltonian to protect the quantum Zeno effect are
established.

A two-level system is then investigated as an example to illustrate
the general results. The decay rate of the survival probability is
derived with and without the coherent control scheme respectively,
and the results indicate that the coherent quantum control scheme
performs well in lowering the decay rate in the presence of dephasing
and amplitude damping noise. As different control Hamiltonians lead
to different suppression effects on the decay of survival probability,
the control Hamiltonian is further optimized to minimize the decay
rate. An optimization equation for the control Hamiltonian is formally
obtained by a variational method and solved analytically for the two
types of noise respectively. The mechanism of how the optimal control
Hamiltonian protects the system against the noise and mitigates the
decay of survival probability is numerically illustrated by visualizing
the noisy evolution paths of the quantum system in the Bloch sphere.
The results show that the effect of the optimal Hamiltonian control
is to rotate the system towards the direction that is least influenced
by the noise and then turns it back to the vicinity of the initial
state, so that the final state with the optimal control can be closer
to the initial state than without the control, and thus the survival
probability of the system to stay in the initial state can be increased.

We hope this work can contribute a novel quantum control strategy
to mitigate the influence of Markovian noise on the quantum Zeno effect
and stimulate future research in this direction.
\begin{acknowledgments}
This work is supported by the National Natural Science Foundation
of China (Grant No. 12075323) and the Innovation Program for Quantum
Science and Technology (Grant No. 2021ZD0300702).
\end{acknowledgments}

\bibliographystyle{apsrev4-2}
\bibliography{reference}

\appendix
\onecolumngrid

\section{Derivation of representation transformation of superoperators in
liouville space\label{sec:Derivation-of-the-1}}

In this appendix, we briefly demonstrate how the transformations of
superoperators $\huaLh[0]$and $\huaL[\inten]$ are derived in Eq.
\eqref{eq:ele}.

According to the definition of $\huaLh[0]$, $\huaLh[0]=-i[H_{0},\cdot]$,
and the definition of $e^{\omega\huaLh[\control]\eta}$, $e^{\omega\huaLh[\control]\eta}\left[\cdot\right]=e^{-i\omega\hami[\control]\eta}\left(\cdot\right)e^{i\omega\hami[\control]\eta}$,
the transformation of the commutator $\huaLh[0]$ into a representation
rotated by $e^{-\omega\huaLh[\control]\eta}$ is
\begin{equation}
\begin{aligned}e^{-\omega\huaLh[\control]\eta}\huaLh[0]e^{\omega\huaLh[\control]\eta}\left[\cdot\right] & =-ie^{i\omega\hami[\control]\eta}\left[\hami[0]e^{-i\omega\hami[\control]\eta}\left(\cdot\right)e^{i\omega\hami[\control]\eta}-e^{-i\omega\hami[\control]\eta}\left(\cdot\right)e^{i\omega\hami[\control]\eta}\hami[0]\right]e^{-i\omega\hami[\control]\eta}\\
 & =-i\left[e^{i\omega\hami[\control]\eta}\hami[0]e^{-i\omega\hami[\control]\eta},\cdot\right]\\
 & =-i\left[\widetilde{H}_{0}(\eta),\cdot\right]=\widetilde{\mathcal{L}}_{H_{\mathrm{0}}}^{(\eta)},
\end{aligned}
\label{eq:transh0}
\end{equation}
where $\widetilde{H}_{0}(\eta)=e^{i\omega\hami[\control]\eta}\hami[0]e^{-i\omega\hami[\control]\eta}$
is the transformed free Hamiltonian in the rotated representation
dependent on the parameter $\eta$, and the transformation of the
dissipative superoperator $\huaL[\inten]$ is
\begin{equation}
\begin{aligned}e^{-\omega\huaLh[\control]\eta}\huaL[\inten]e^{\omega\huaLh[\control]\eta}\left[\cdot\right] & =\sum_{k}\inten_{k}e^{-\omega\huaLh[\control]\eta}\huaD[V_{k}]e^{\omega\huaLh[\control]\eta}\left(\cdot\right)\\
 & =\sum_{k}\inten_{k}e^{i\omega\hami[\control]\eta}\big\{ V_{k}\left[e^{-i\omega\hami[\control]\eta}\left(\cdot\right)e^{i\omega\hami[\control]\eta}\right]V_{k}^{\dagger}\\
 & -\frac{1}{2}V_{k}^{\dagger}V_{k}\left[e^{-i\omega\hami[\control]\eta}\left(\cdot\right)e^{i\omega\hami[\control]\eta}\right]-\frac{1}{2}\left[e^{-i\omega\hami[\control]\eta}\left(\cdot\right)e^{i\omega\hami[\control]\eta}\right]V_{k}^{\dagger}V_{k}\big\} e^{-i\omega\hami[\control]\eta}\\
 & =\sum_{k}\inten_{k}\left[\widetilde{V_{k}}(\eta)\left(\cdot\right)\widetilde{V_{k}}^{\dagger}(\eta)-\frac{1}{2}\left\{ \widetilde{V_{k}}^{\dagger}(\eta)\widetilde{V_{k}}(\eta),\cdot\right\} \right]\\
 & =\sum_{k}\inten_{k}\huaD[\widetilde{V_{k}}(\eta)],
\end{aligned}
\label{eq:transmu}
\end{equation}
where $\widetilde{V_{k}}(\eta)=e^{i\omega\hami[\control]\eta}V_{k}e^{-i\omega\hami[\control]\eta}$
is the transformed noise operator $V_{k}$ in the framework rotated
by $e^{-i\omega\hami[\control]\eta}$ dependent on the parameter $\eta$.

It can be observed that the transformations of the superoperators
$\huaLh[0]$and $\huaL[\inten]$ are essentially the transformations
of the free Hamiltonian $H_{0}$ and the noise operators $V_{k}$
under the control Hamiltonian $\hami[\control]$, respectively, while
the forms of $\huaLh[0]$and $\huaL[\inten]$, regardless of $H_{0}$
and $V_{k}$, remain unchanged.

\section{Derivation of optimization equation for control Hamiltonian\label{subsec:two-level-quantum-system}}

In this appendix, we derive the equation that determine the optimal
control Hamiltonians to minimize the effective decay rates of survival
probability $\gammaeff^{(\control)}$ for a two-level quantum system.%

We start from the general result for $\gammaeff^{(\control)}$,
\begin{equation}
\begin{aligned}\gammaeff^{(\control)} & =-\frac{3}{2}(\boldsymbol{n_{\control}}\cdot\boldsymbol{r_{0}})^{2}\boldsymbol{n_{\control}}\Gamma\boldsymbol{n_{\control}}+\frac{1}{2}\boldsymbol{n_{\control}}\cdot\boldsymbol{r_{0}}(\boldsymbol{n_{\control}}\Gamma\boldsymbol{r_{0}}+\boldsymbol{r_{0}}\Gamma\boldsymbol{n_{\control}})-\frac{1}{2}\boldsymbol{r_{0}}\Gamma\boldsymbol{r_{0}}\\
 & -\frac{1}{2}\left(\boldsymbol{n_{\control}}\times\boldsymbol{r_{0}}\right)\Gamma\left(\boldsymbol{n_{\control}}\times\boldsymbol{r_{0}}\right)+\tr\Gamma+(\boldsymbol{n_{\control}}\cdot\boldsymbol{r_{0}})(\mathbf{g\cdot n_{c}}),
\end{aligned}
\end{equation}
which is provided by Eq. \eqref{eq:66} in Subsec. \ref{subsec:Scheme-Introduced-Channel}.

Denote the Bloch vector of the initial state as $\boldsymbol{r_{0}}$
and the normalized directional vector of the control Hamiltonian as
$\boldsymbol{n_{\control}}$, i.e., $H_{\control}=\boldsymbol{n_{\control}}\cdot\vp$.
The cross product between the vectors $\boldsymbol{n_{\control}}$
and $\boldsymbol{r_{0}}$ can be represented as a linear transformation
of $\boldsymbol{n_{\control}}$, given by
\begin{equation}
\boldsymbol{n_{\control}}\times\boldsymbol{r_{0}}=R\boldsymbol{n_{\control}}.
\end{equation}
If $\boldsymbol{r_{0}}$ is denoted as
\begin{equation}
\boldsymbol{r_{0}}=(x_{0},y_{0},z_{0}),
\end{equation}
the transformation matrix $R$ is antisymmetric and defined as
\begin{equation}
R=\begin{bmatrix}0 & z_{0} & -y_{0}\\
-z_{0} & 0 & x_{0}\\
y_{0} & -x_{0} & 0
\end{bmatrix}.
\end{equation}

In this case, the effective decay rate of survival probability $\gammaeff^{(\control)}$
can be rewritten as
\begin{equation}
\begin{aligned}\gammaeff^{(\control)} & =-\frac{3}{2}(\boldsymbol{r_{0}^{\zhuan}}\boldsymbol{n_{\control}})(\boldsymbol{n_{\control}^{\zhuan}}\Gamma\boldsymbol{n_{\control}})(\boldsymbol{n_{\control}^{\zhuan}}\boldsymbol{r_{0}})+\frac{1}{2}(\boldsymbol{r_{0}^{\zhuan}}\boldsymbol{n_{\control}})(\boldsymbol{n_{\control}^{\zhuan}}\Gamma\boldsymbol{r_{0}})+\frac{1}{2}(\boldsymbol{r_{0}^{\zhuan}}\Gamma\boldsymbol{n_{\control}})(\boldsymbol{n_{\control}^{\zhuan}}\boldsymbol{r_{0}})-\frac{1}{2}\boldsymbol{r_{0}^{\zhuan}}\Gamma\boldsymbol{r_{0}}\\
 & -\frac{1}{2}\left(R\boldsymbol{n_{\control}}\right)^{\zhuan}\Gamma\left(R\boldsymbol{n_{\control}}\right)+\tr\Gamma+(\nb^{\zhuan}\boldsymbol{n_{\control}})(\boldsymbol{n_{\control}^{\zhuan}}\boldsymbol{r_{0}})\\
 & =\tr\left(-\frac{3}{2}P_{\zheng n}\Gamma P_{\zheng n}P_{\zheng r}+\frac{1}{2}P_{\zheng n}\Gamma P_{\zheng r}+\frac{1}{2}\Gamma P_{\zheng n}P_{\zheng r}-\frac{1}{2}\Gamma P_{\zheng r}-\frac{1}{2}R^{\zhuan}\Gamma RP_{\zheng n}+\Gamma+P_{\zheng n}\boldsymbol{r_{0}}\nb^{\zhuan}\right),
\end{aligned}
\end{equation}
where $P_{\zheng n}$ and $P_{\zheng r}$ are defined as $P_{\zheng n}=\boldsymbol{n_{\control}}\boldsymbol{n_{\control}^{\zhuan}}$
and $P_{\zheng r}=\boldsymbol{r_{0}}\boldsymbol{r_{0}^{\zhuan}}$
respectively, and the superscript ``$\zhuan$'' denotes the transposition
of a column vector.

Considering the normalization of the vector $\boldsymbol{n_{\control}}$,
i.e., $\left\Vert \boldsymbol{n_{\control}}\right\Vert =1$, $P_{\zheng n}$
is actually a projection operator, satisfying $P_{\zheng n}^{2}=P_{\zheng n}$,
so the Lagrangian function should include this property of $P_{\zheng n}$
as a constraint condition,
\begin{equation}
L\left(P_{\zheng n},\Lambda\right)=\gammaeff^{(\control)}+\tr\left[\left(P_{\zheng n}^{2}-P_{\zheng n}\right)\Lambda\right],\label{eq:lagr}
\end{equation}
where $\Lambda$ is an arbitrary matrix, representing the Lagrange
multiplier.

To obtain the optimal control Hamiltonian $\hami[\control]$ that
minimizes the effective decay rate $\gammaeff^{(\control)}$, we perform
variational calculus on the Lagrangian function \eqref{eq:lagr},
yielding
\begin{equation}
\begin{aligned}\delta L & =\tr\left[\delta P_{\zheng n}\left(-\frac{3}{2}\Gamma P_{\zheng n}P_{\zheng r}-\frac{3}{2}P_{\zheng r}P_{\zheng n}\Gamma+\frac{1}{2}\Gamma P_{\zheng r}+\frac{1}{2}P_{\zheng r}\Gamma-\frac{1}{2}R^{\zhuan}\Gamma R+\boldsymbol{r_{0}}\nb^{\zhuan}+P_{\zheng n}\Lambda+\Lambda P_{\zheng n}-\Lambda\right)\right]\\
 & +\tr\left[\delta\Lambda\left(P_{\zheng n}^{2}-P_{\zheng n}\right)\right].
\end{aligned}
\label{eq:deltaL}
\end{equation}

Accroding to the principle of the variational approach, the variation
$\delta L$ in Eq. \eqref{eq:deltaL} should be zero for any $\delta P_{\zheng n}$
and $\delta\Lambda$, leading to the following conditions for minimizing
the effective decay rate $\gammaeff^{(\control)}$:
\begin{align}
-\frac{3}{2}\Gamma P_{\zheng n}P_{\zheng r}-\frac{3}{2}P_{\zheng r}P_{\zheng n}\Gamma+\frac{1}{2}\Gamma P_{\zheng r}+\frac{1}{2}P_{\zheng r}\Gamma-\frac{1}{2}R^{\zhuan}\Gamma R+\boldsymbol{r_{0}}\nb^{\zhuan}+P_{\zheng n}\Lambda+\Lambda P_{\zheng n}-\Lambda=\mathbf{0},\\
P_{\zheng n}^{2}-P_{\zheng n}=\mathbf{0},
\end{align}
where the bold symbol $\boldsymbol{0}$ denotes the zero matrix.

\section{Derivation of effective decay rate of surviaval probability for two-level
system\label{sec:Derivation-of-effective}}

This appendix focuses primarily on the coherent control scheme for
two-level system, specifically the derivations discussed in Sec.$\ $\ref{sec:STRONG-QUANTUM-CONTROL-1}.
We start from the general results of effective decay rates $\gammaeff^{(\noise)}$
\eqref{eq:22} without quantum control and $\gammaeff^{(\control)}$
\eqref{eq:refc} with coherent quantum controls in Sec. \ref{sec:STRONG-QUANTUM-CONTROL},
and apply them to a two-level system in the presence of Markovian
noise with and without the coherent control scheme respectively.

The dissipative superoperator $\huaL[\mu]$ induced by Markovian noise
for a two-level system can be generally expressed as
\begin{equation}
\huaL[\mu]\left[\cdot\right]=\sum_{ij}\inten_{ij}\left(\sigma_{i}\left(\cdot\right)\sigma_{j}-\frac{1}{2}\left\{ \sigma_{j}\sigma_{i},\cdot\right\} \right),\label{eq:twolevelL}
\end{equation}
which is given in Eq. \eqref{eq:47}. By substituting this equation
into $\gammaeff^{(\noise)}$ \eqref{eq:22}, one can derive the effective
decay rate for a noisy two-level quantum system without quantum control
as
\begin{equation}
\begin{aligned}\gammaeff^{(\noise)} & =-\lef[\psio]\huaL[\mu]\left[\rou[0]\right]\rig[\psio]\\
 & =-\sum_{i,j=1,2}\inten_{ij}\ \tr\left(\rou[0]\sigma_{i}\rou[0]\sigma_{j}-\frac{1}{2}\rou[0]\left\{ \sigma_{j}\sigma_{i},\rou[0]\right\} \right)\\
 & =-\sum_{i,j=1,2}\inten_{ij}\left[\tr\left(\sigma_{i}\rou[0]\right)\tr\left(\sigma_{j}\rou[0]\right)-\tr\left(\sigma_{j}\sigma_{i}\rou[0]\right)\right]\\
 & =-\boldsymbol{r_{0}^{\zhuan}}\Gamma\boldsymbol{r_{0}}+\tr\Gamma-i\sum_{ijk}\inten_{ij}\varepsilon_{ijk}\left(r_{0}\right)_{k}\\
 & =-\boldsymbol{r_{0}^{\zhuan}}\Gamma\boldsymbol{r_{0}}+\tr\Gamma+\nb\cdot\boldsymbol{r_{0}}
\end{aligned}
,\label{eq:75}
\end{equation}
where the third line of the derivation results from the assumption
that $\rou[0]$ is the density matrix of a pure state\textbf{, $\nb$}
is a vector related to the imaginary parts of the off-diagonal elements
of the noise coefficient matrix $\Gamma$, defined as
\begin{equation}
\nb=2(\im\mu_{23},\,\im\mu_{31},\,\im\mu_{12}).\label{eq:g-1}
\end{equation}

In the coherent control scheme, we assume that the control Hamiltonian
can be written as
\begin{equation}
\hami[\control]=\boldsymbol{n_{\control}}\cdot\vp,\label{eq:hc}
\end{equation}
where $\boldsymbol{n_{\control}}$ describes the direction of the
control Hamiltonian,
\begin{equation}
\boldsymbol{n_{\control}}=(\sin\thetakong[\control]\cos\ph[\control],\sin\thetakong[\control]\sin\ph[\control],\cos\thetakong[\control]).
\end{equation}
The eigenvalues and the associated eigenstates of the control Hamiltonian
$\hami[\control]$ can be straightforwardly obtained as
\begin{equation}
\begin{cases}
E_{1}^{(\control)}=1, & |\psi_{1}^{(\control)}\rangle=e^{-i\ph[\control]}\cos\frac{\thetakong[\control]}{2}|0\rangle+\sin\frac{\thetakong[\control]}{2}|1\rangle\\
E_{2}^{(\control)}=-1, & |\psi_{2}^{(\control)}\rangle=-e^{-i\ph[\control]}\sin\frac{\thetakong[\control]}{2}|0\rangle+\cos\frac{\thetakong[\control]}{2}|1\rangle
\end{cases},\label{eq:50}
\end{equation}
where $E_{1}^{(\control)}$, $E_{2}^{(\control)}$ are the eigenvalues
and $|\psi_{1}^{(\control)}\rangle$, $|\psi_{2}^{(\control)}\rangle$
are the eigenstates.

Correspondingly, an arbitrary initial state $\rig[\psio]$ in the
basis of the Hamiltonian's eigenstates can be expressed as 
\begin{equation}
\rig[\psio]=a_{\zheng 1}|\psi_{1}^{(\control)}\rangle+a_{\zheng 2}|\psi_{2}^{(\control)}\rangle,\label{eq:63}
\end{equation}
where the coefficients $a_{\zheng 1}$, $a_{\zheng 2}$ satisfy the
normalization condition $\abs{a_{1}}^{2}+\abs{a_{2}}^{2}=1$.

Now, let us derive the effective decay rate $\gammaeff^{(\control)}$
for a two-level system within scheme with the coherent quantum control.
The first consideration pertains to the necessary conditions for the
validity of the coherent control scheme in a two-level system. Substituting
the eigenvalues from Eq. \eqref{eq:50} into the zeroth-order term
of survival probability $\prot[\control](\tau)$ in Eq. \eqref{eq:59},
one can have
\begin{equation}
\prot[\control]^{\zheng{}}|_{\tau=0}=\abs{\lef[\psio]e^{i\omega\hami[\control]}\rig[\psio]}^{2}=|a_{1}|^{4}+|a_{2}|^{4}+2|a_{1}|^{2}|a_{2}|^{2}\cos2\omega,
\end{equation}
indicating that when $\omega=n\pi$ for $n=\pm1,\pm2,\pm3\cdots$,
$\prot[\control]^{\zheng{}}|_{\tau=0}=1$ by the normalization condition
of $a_{\zheng 1}$, $a_{\zheng 2}$.

With the specific form of $\huaL[\mu]$ for a two-level system given
in Eq. \eqref{eq:twolevelL} and the representation transformation
of superoperator $\huaL[\inten]$, $\widetilde{\mathcal{L}}_{\zheng{\inten}}^{\zheng{(\eta)}}=e^{-\omega\huaLh[\control]\eta}\huaL[\inten]e^{\omega\huaLh[\control]\eta}$,
in Eq. \eqref{eq:transmu}, the first-order term of Eq. \eqref{eq:59}
can be written as
\begin{equation}
\prot[c]^{\zheng{(1)}}|_{\tau=0}=\int_{0}^{1}\lef[\psio]\widetilde{\mathcal{L}}_{\zheng{\inten}}^{\zheng{(\eta)}}\left[\rou[0]\right]\rig[\psio]d\eta=\sum_{i,j=1}^{3}\inten_{ij}\int_{0}^{1}\tr\left[\rou[0]\widetilde{\sigma}_{i}^{(\eta)}\rou[0]\widetilde{\sigma}_{j}^{(\eta)}-\rou[0]\widetilde{\sigma}_{j}^{(\eta)}\widetilde{\sigma}_{i}^{(\eta)}\right]d\eta,\label{eq:pc1}
\end{equation}
where $\widetilde{\sigma}_{i}^{(\eta)}\equiv e^{i\omega\hami[\control]\eta}\sigma_{i}e^{-i\omega\hami[\control]\eta}$
denotes the Pauli operator $\sigma_{i}$ in the transformed framework.
As $H_{\control}$ is a normalized Pauli matrix defined in \eqref{eq:hc},
satisfying $H_{\control}^{2}=I$, one can have
\begin{equation}
e^{\pm i\omega\hami[\control]\eta}=\cos\left(\omega\eta\right)\mathbb{I}\pm i\sin\left(\omega\eta\right)H_{\control}.\label{eq:50-1}
\end{equation}

Considering the necessary condition $\omega=n\pi,$ $n=\pm1,\pm2,\pm3\cdots$,
substituting above equation, $H_{\control}=\boldsymbol{n_{\control}}\cdot\vp$
and $\rho_{0}=\left(\mathbb{I}+\boldsymbol{r_{0}}\cdot\vp\right)/2$
into Eq.$\ $\eqref{eq:pc1} and using the identity
\begin{equation}
(\boldsymbol{a}\cdot\boldsymbol{\sigma})(\boldsymbol{b}\cdot\boldsymbol{\sigma})=(\boldsymbol{a}\cdot\boldsymbol{b})\ \mathbb{I}+i\left(\boldsymbol{a}\times\boldsymbol{b}\right)\cdot\boldsymbol{\sigma},
\end{equation}
for two arbitrary vectors $\boldsymbol{a}$ and $\boldsymbol{b}$,
one can obtain
\begin{align}
p_{\control}^{\left(1\right)}|_{\tau=0} & =\sum_{i,j=1}^{3}\inten_{ij}\intop_{0}^{1}\tr\left(\rou[0]e^{i\omega\hami[\control]\eta}\sigma_{i}e^{-i\omega\hami[\control]\eta}\rou[0]e^{i\omega\hami[\control]\eta}\sigma_{j}e^{-i\omega\hami[\control]\eta}-\rou[0]e^{i\omega\hami[\control]\eta}\sigma_{j}\sigma_{i}e^{-i\omega\hami[\control]\eta}\right)d\eta\\
 & =\sum_{i,j=1}^{3}\inten_{ij}\bigg\{\intop_{0}^{1}\tr\big(\rou[0]\left[\cos\left(\omega\eta\right)\ \mathbb{I}+i\sin\left(\omega\eta\right)\ \boldsymbol{n_{\control}}\cdot\vp\right]\sigma_{i}\left[\cos\left(\omega\eta\right)\ \mathbb{I}-i\sin\left(\omega\eta\right)\ \boldsymbol{n_{\control}}\cdot\vp\right]\rou[0]\nonumber \\
 & \ \ \left[\cos\left(\omega\eta\right)\ \mathbb{I}+i\sin\left(\omega\eta\right)\ \boldsymbol{n_{\control}}\cdot\vp\right]\sigma_{j}\left[\cos\left(\omega\eta\right)\ \mathbb{I}+i\sin\left(\omega\eta\right)\ \boldsymbol{n_{\control}}\cdot\vp\right]d\eta\big)\nonumber \\
 & \ \ -\intop_{0}^{1}\tr\big(\rou[0]\left[\cos\left(\omega\eta\right)\ \mathbb{I}+i\sin\left(\omega\eta\right)\ \boldsymbol{n_{\control}}\cdot\vp\right]\sigma_{j}\sigma_{i}\left[\cos\left(\omega\eta\right)\ \mathbb{I}-i\sin\left(\omega\eta\right)\ \boldsymbol{n_{\control}}\cdot\vp\right]d\eta\big)\bigg\}\\
 & =\sum_{i,j=1}^{3}\inten_{ij}\bigg\{\frac{3}{2}\left(\boldsymbol{n_{\control}}\cdot\boldsymbol{r_{0}}\right)^{2}\left(\boldsymbol{n_{\control}}\right)_{i}\left(\boldsymbol{n_{\control}}\right)_{j}-\frac{1}{2}\boldsymbol{n_{\control}}\cdot\boldsymbol{r_{0}}\left[\left(\boldsymbol{n_{\control}}\right)_{i}\left(\boldsymbol{r_{0}}\right)_{j}+\left(\boldsymbol{r_{0}}\right)_{i}\left(\boldsymbol{n_{\control}}\right)_{j}\right]+\frac{1}{2}\left(\boldsymbol{r_{0}}\right)_{i}\left(\boldsymbol{r_{0}}\right)_{j}\nonumber \\
 & \ \ +\frac{1}{2}\left(\boldsymbol{n_{\control}}\times\boldsymbol{r_{0}}\right)_{i}\left(\boldsymbol{n_{\control}}\times\boldsymbol{r_{0}}\right)_{j}-\delta_{ij}+i(\boldsymbol{n_{\control}}\cdot\boldsymbol{r_{0}})\sum_{k}\varepsilon_{ijk}\left(\boldsymbol{n_{\control}}\right)_{k}\bigg\},
\end{align}
where $\left(\boldsymbol{n_{\control}}\right)_{k}$ , $\left(\boldsymbol{r_{0}}\right)_{k}$
and $\left(\boldsymbol{n_{\control}}\times\boldsymbol{r_{0}}\right)_{k}$
denote the $k$-th elements of vectors $\boldsymbol{n_{\control}}$,
$\boldsymbol{r_{0}}$ and $\boldsymbol{n_{\control}}\times\boldsymbol{r_{0}}$,
respectively.

Substituting the real vector $\nb$ defined in Eq.$\ $\eqref{eq:g-1}
into the above equation, one can finally arrive at
\begin{align}
p_{\control}^{\left(1\right)}|_{\tau=0} & =\frac{3}{2}\left(\boldsymbol{n_{\control}}\cdot\boldsymbol{r_{0}}\right)^{2}\boldsymbol{n_{\control}^{\zhuan}}\Gamma\boldsymbol{n_{\control}}-\frac{1}{2}(\boldsymbol{n_{\control}}\cdot\boldsymbol{r_{0}})\left(\boldsymbol{n_{\control}^{\zhuan}}\Gamma\boldsymbol{r_{0}}+\boldsymbol{r_{0}^{\zhuan}}\Gamma\boldsymbol{n_{\control}}\right)+\frac{1}{2}\boldsymbol{r_{0}^{\zhuan}}\Gamma\boldsymbol{r_{0}}\nonumber \\
 & \ \ +\frac{1}{2}\left(\boldsymbol{n_{\control}}\times\boldsymbol{r_{0}}\right)^{\boldsymbol{{\rm T}}}\Gamma\left(\boldsymbol{n_{\control}}\times\boldsymbol{r_{0}}\right)-\tr\Gamma-(\boldsymbol{n_{\control}}\cdot\boldsymbol{r_{0}})(\nb\cdot\boldsymbol{n_{\control}}),
\end{align}
which is Eq. \eqref{eq:66} in the main text.
\end{document}